\documentclass{aa}  
\usepackage{graphicx}
\usepackage{longtable}
\usepackage{txfonts}

\usepackage{color}
\begin{document}

   \title{Astrophysical parameters and orbital solution of the peculiar X-ray transient IGR~J00370+6122}

   \author{A. Gonz\'alez-Gal\'an
          \inst{1}
          \and
          I. Negueruela \inst{1}
          \and
          N. Castro \inst{2}
          \and
          S. Sim\'on-D\'iaz \inst{3,4}
          \and 
          J. Lorenzo \inst{1}
          \and
          F. Vilardell \inst{1,5}
                }
          
   \institute{Departamento de F\'isica, Ingenier\'ia de Sistemas y Teor\'ia de la Se\~nal. Universidad de Alicante, Apdo. 99, 03080 Alicante, Spain\\
              \email{anagonzalez@ua.es}
           \and
                Argelander Institut f\"ur Astronomie, Auf den H\"ugel 71, Bonn, 53121, Germany 
           \and 
                Instituto de Astrof\'isica de Canarias, V\'ia L\'actea s/n, E-38205 La Laguna, Santa Cruz de Tenerife, Spain
           \and 
                Departamento de Astrof\'isica, Facultad de F\'isica y Matem\'aticas, Universidad de La Laguna, Avda. Astrof\'isico Francisco S\'anchez, s/n, E-38206 La Laguna, Santa Cruz de Tenerife, Spain
 \and
                Institut d'Estudis Espacials de Catalunya, Edifici Nexus, c/ Gran Capit\`a, 2-4, desp. 201, E-08034 Barcelona, Spain}

\titlerunning{Parameters and orbit for IGR~J00370+6122}

  \abstract
   { BD~$+60^\circ\,$73 is the optical counterpart of the X-ray source IGR~J00370+6122, a probable accretion-powered X-ray pulsar. The X-ray light curve of this binary system shows clear periodicity at 15.7~d, which has been interpreted as repeated outbursts around the periastron of an eccentric orbit.}
   {We aim to characterise the binary system IGR~J00370+6122 by deriving its orbital and physical parameters.}
   {We obtained high-resolution spectra of BD~$+60^\circ\,$73 at different epochs. We used the \textsc{fastwind} code to generate a stellar atmosphere model to fit the observed spectrum and obtain physical magnitudes. The synthetic spectrum was used as a template for cross-correlation with the observed spectra to measure radial velocities. The radial velocity curve provided an orbital solution for the system. We also analysed the \textit{RXTE}/ASM and \textit{Swift}/BAT light curves to confirm the stability of the periodicity.}
   {BD~$+60^\circ\,$73 is a BN0.7\,Ib low-luminosity supergiant located at a distance $\sim$3.1~kpc, in the Cas~OB4 association. We derive $T_{{\rm eff}}=24\,000$~K and $\log g_{\rm c}=3.0$, and chemical abundances consistent with a moderately high level of evolution. The spectroscopic and evolutionary masses are consistent at the 1-$\sigma$ level with a mass $M_{*}\approx15\,M_{\sun}$. The recurrence time of the X-ray flares is the orbital period of the system. The neutron star is in a high-eccentricity ($e=0.56\pm0.07$) orbit, and the X-ray emission is strongly peaked around orbital phase $\phi=0.2$, though the observations are consistent with some level of X-ray activity happening at all orbital phases.}
   {The X-ray behaviour of IGR~J00370+6122 is reminiscent of ``intermediate'' supergiant X-ray transients, though its peak luminosity is rather low. The orbit is somewhat wider than those of classical persistent supergiant X-ray binaries, which when combined with the low luminosity of the mass donor,  explains the low X-ray luminosity. IGR~J00370+6122 will very likely evolve towards a persistent supergiant system, highlighting the evolutionary connection between different classes of wind-accreting X-ray sources.}

   \keywords{stars: binaries: close - stars: evolution - stars: individual: BD +60$^{\circ}$73  - stars: pulsars - stars: supergiants - X-rays: stars - X-rays: individual: IGR J00370+6122
               }
 \date{Received; accepted}
   \maketitle
%

\section{Introduction}

X-ray binaries (XRBs) are systems that consist of a compact object (neutron star or a black hole) orbiting an optical companion. High-mass X-ray binaries (HMXBs) are a subclass of XRBs where the optical companion is an early-type (O or B-type) star with a mass above $\sim$8$\:M_{\sun}$, i.e. a high-mass star. HMXBs are strong emitters of X-ray radiation as a result of accretion of matter from the OB companion onto the compact object. Based on the type of the optical star, HMXBs are classified further into supergiant X-ray binaries (SGXBs) and Be/X-ray binaries (BeXBs).

In BeXBs, material from a circumstellar disc is accreted onto the compact object. Most BeXBs have relatively wide orbits ($P_{{\rm orb}}\ga 20$~d), with moderate to high eccentricities \citep[$e\gtrsim0.3$;][]{reig2011}, and the compact companion spends most of its time far away from the disc surrounding the Be star \citep{heuvel1987}. In consequence, they are transient sources, with the outbursts often occurring at regular intervals, separated by the orbital period, and generally close to periastron \citep{negueruela2001}. Persistent BeXBs also exist \citep{reig1999}, showing much less X-ray variability and lower luminosities \citep[$L_{\rm X}$$\lesssim$$10^{35}$~erg~s$^{-1}$;][]{reig2011}. 

In SGXBs the optical companion is an OB supergiant. Supergiant stars are known to suffer great mass loss in the form of a radiation-driven stellar wind \citep{kudritzki00}. The compact star interacts with this wind, producing the persistent X-rays observed in SGXBs \citep[with typically $L_{\rm X}\sim10^{36}$~erg~s$^{-1}$;][]{nagase1989} with occasional flaring variability on short time scales (seconds), but rather stable in the long run. Almost all these systems are X-ray pulsars, and orbital solutions exist for most of them. They have orbital periods ranging from $\gtrsim3$~d to $\sim15$~d and present circular or low-eccentricity orbits \citep[e.g.][]{negueruela2008}, with the single exception of GX~301$-$2, powered by a B1.5\,Ia$^{+}$ hypergiant. Many of them display eclipses of the X-ray source.

One relatively recent discovery in this field is the existence of systems having X-ray fast transient phenomena, generally with a steeper rise and slower decay of the X-ray flux of several orders of magnitude, and lasting from minutes up to a few hours \citep{sguera2006}. Fast transients have always been identified with supergiant companions \citep[e.g. ][]{zand2005,negueruela2006a,blay2012}, leading to the definition of the supergiant fast X-ray transients \citep[SFXTs,][]{negueruela2006b} as a sub-class of SGXBs. SFXTs spend most of their time at X-ray luminosities between $L_{\rm X}\sim10^{32}$ to $10^{34}$~erg~s$^{-1}$, well below the "normal" X-ray luminosity of SGXBs, with very brief excursions to outburst luminosities of up to a few times $10^{36}$~erg~s$^{-1}$ \citep[e.g.][]{rampy2009,sidoli2009,sidoli12}. The underlying mechanism that produces the fast X-ray outburst in SFXTs is still not understood well.

A number of hypothesis have been put forward to explain the different behaviours in SFXTs and "normal" SGXBs. Some of the proposed models invoke the structure in the wind of the supergiant companion that could produce sudden episodes of accretion onto the compact component. This structure could be either in the form of clumping in a spherically symmetric outflow from the supergiant donor \citep{zand2005,walter2007,negueruela2008,zurita2009,blay2012} or in the form of an equatorial density enhancement in the wind from the supergiant, inclined at some angle to the orbit of the neutron star \citep{romano2007,sidoli2007}. Another possible explanation is that there is an eccentric orbit that could lead the neutron star to cross zones of large and variable absorption \citep{negueruela2008, blay2012}. Of course, a clumpy spherical wind can exist together with an eccentric orbit \citep{blay2012}, and this scenario would also explain the quasi-stable X-ray luminosity of SGXBs, since they have circular orbits. Furthermore, \citet{bozzo2008} proposed a model that makes use of transitions between accretion gating mechanisms, such as centrifugal and magnetic barriers, brought about by variations in the stellar wind, to explain the wide dynamic range in flux observed in this systems. \cite{ducci2010} propose a scenario that links all these mechanisms.

IGR~J00370+6122 is a HMXB discovered in December 2003 during a 1.2~Ms INTEGRAL observation \citep{hartog2004,hartog2006}. The error circle of this detection included the {\it ROSAT} source 1RXS~J00709.6+612131, which had been identified with the OB star BD~$+60^\circ\,$73 = LS~I~$+61^\circ$161 \citep{rutled2000,rutled2004}. This source was originally classified as a B1\,Ib star by \cite{morgan55}. Based on an intermediate-resolution spectrum, \citet{reig2005} classified BD~$+60^\circ\,$73 as a BN0.5\,II--III star, without any evidence of a circumstellar disc. This spectral type would place IGR~J00370+6122 outside the two main divisions of HMXBs. (It is neither a Be star nor a supergiant.)

Using data from the All-Sky Monitor (ASM) on {\em RXTE}, \citet{hartog2004} find that the X-ray flux from IGR~J00370+6122 is very strongly modulated at $15.665\pm0.006$~d. \citet{zand2007} interpret its behaviour as a series of outbursts separated by the orbital period. The light curve is dominated by fast flaring and the absorption column is much higher than the interstellar value, leading to the suggestion that the system is powered by wind accretion in a very eccentric orbit \citep{zand2007}. 
Pointed {\em RXTE}/PCA observations detected a  very likely modulation at 346$\pm$6~s during a bright X-ray flare, which was interpreted as the detection of the pulsar period from a neutron star  \citep{zand2007}. 
\cite{grunhut2014} have recently published an orbital solution that confirms the high eccentricity expected for the system, and suggests a high inclination of the orbit.

In this paper, we present high-resolution spectroscopy of BD~$+60^\circ\,$73 taken over more than three years (Section~\ref{obs}). Using these data, in Sect.~\ref{data}, we derive the astrophysical parameters of the mass donor by using a model atmosphere fit. We also determine most orbital parameters by analysing the radial velocity curve. Finally, we re-analyse the long-term X-ray light curves taking this orbital solution into account. In Sect.~\ref{disc}, we discuss the accuracy of our results and the constraints they impose on the system properties. We then discuss the information that our results provide for the general class of wind-accreting HMXBs. We finish wrapping up with our conclusions.

\section{Observations}
\label{obs}

\subsection{Optical spectroscopy}

Spectra of BD~$+60^\circ\,$73 were taken during different observing campaigns between 2009 and 2013 (see Table~\ref{log} for the observation log). The observations were obtained with the High Efficiency and Resolution Mercator Echelle Spectrograph (HERMES) operated at the 1.2~m Mercator Telescope (La Palma, Spain) and with the high-resolution FIbre-fed Echelle Spectrograph (FIES) attached to the 2.56~m Nordic Optical Telescope (NOT; also at La Palma).

\begin{figure*}
\resizebox{\textwidth}{!}{
\includegraphics[clip, angle=90]{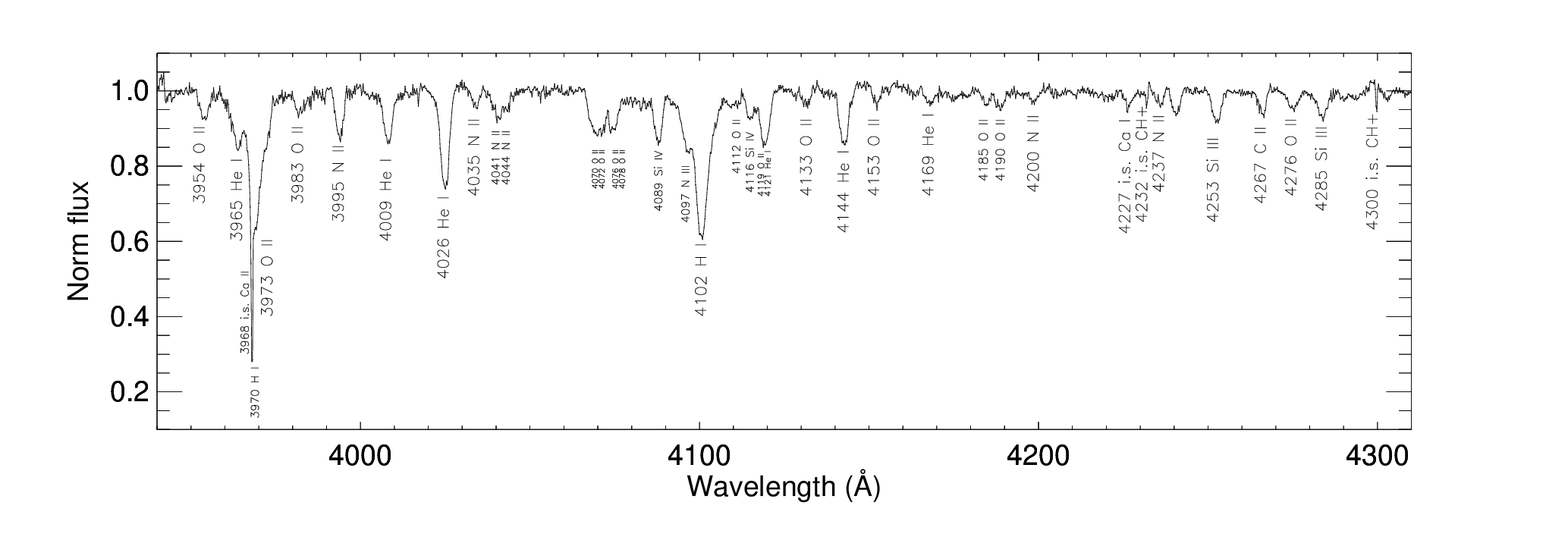}}
\resizebox{\textwidth}{!}{
\includegraphics[clip, angle=90]{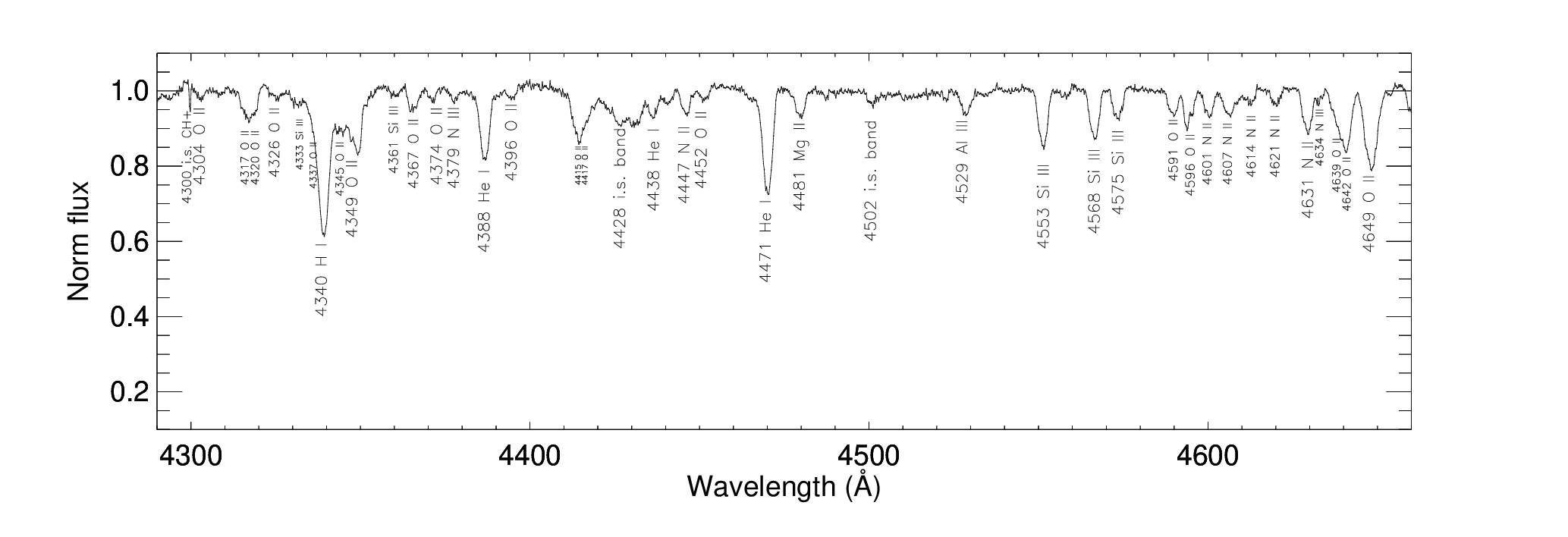}}
\resizebox{\textwidth}{!}{
\includegraphics[clip, angle=90]{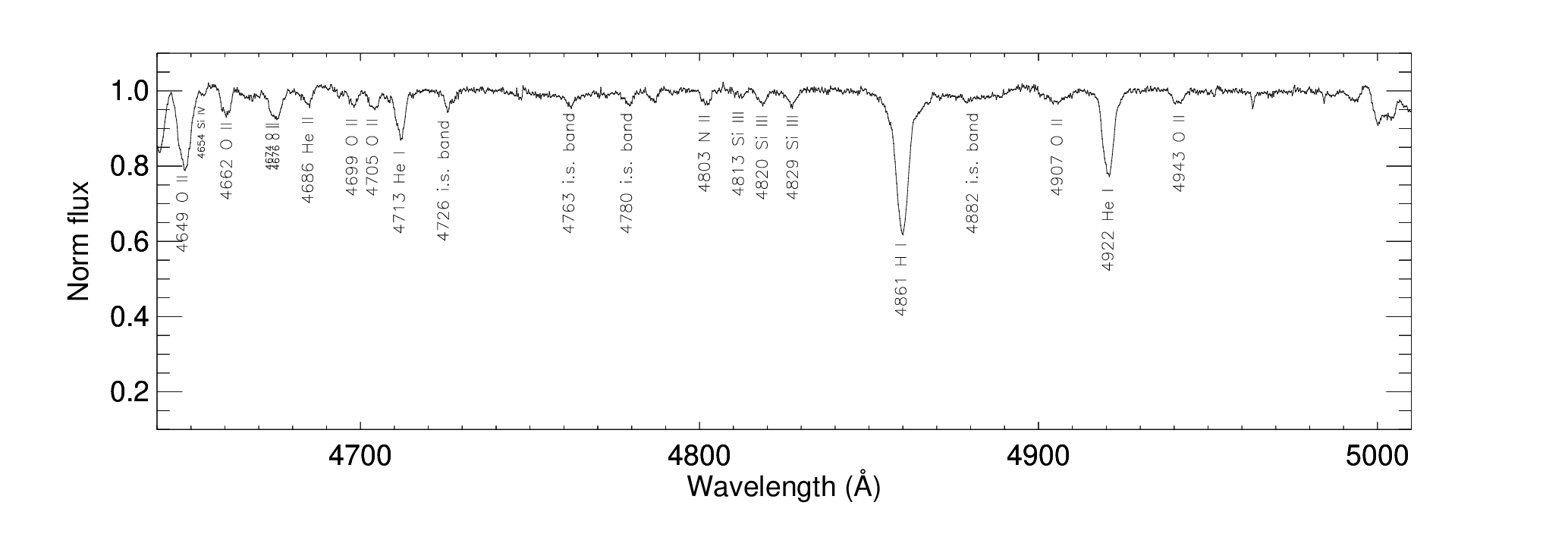}}
\caption{\label{spectrum}Spectrum of BD~$+60^\circ\,$73 over the classification region, obtained with the FIES spectrograph on September 2011, 11.}
\end{figure*}

\begin{table}
\caption{\label{log} Log of high-resolution optical spectra.}
\begin{center}
\begin{tabular}{rccrl}
\hline
\hline
&&&&\\
$\#$ & Date & MJD & $T_{{\rm exp}}$ (s) & Instrument \\
\hline
&&&&\\
1  & \scriptsize{30/10/2009 00:02:06} & 55135.08 & 5400 & HERMES \\
2  & \scriptsize{30/10/2009 23:06:02} & 55136.04 & 5400 & HERMES \\
3  & \scriptsize{31/10/2009 22:27:40} & 55137.02 & 5400 & HERMES \\           
4  & \scriptsize{01/11/2009 21:40:26} & 55137.98 & 7200 & HERMES \\
5  & \scriptsize{02/11/2009 22:15:30} & 55139.01 & 5400 & HERMES \\
6  & \scriptsize{03/11/2009 23:11:47} & 55140.05 & 5400 & HERMES \\
7  & \scriptsize{04/11/2009 21:43:40} & 55140.99 & 5400 & HERMES \\
8  & \scriptsize{06/11/2009 23:34:06} & 55143.06 & 5400 & HERMES \\
9  & \scriptsize{09/11/2009 01:09:17} & 55145.13 & 5400 & HERMES \\
10 & \scriptsize{14/06/2011 04:46:43} & 55726.70 & 1200 & HERMES  \\
11 & \scriptsize{17/06/2011 04:45:41} & 55729.70 & 1200 & HERMES \\
12 & \scriptsize{19/06/2011 05:11:19} & 55731.72 & 1200 & HERMES  \\
13 & \scriptsize{20/06/2011 04:06:44} & 55732.67 & 1200 & HERMES  \\
14 & \scriptsize{11/09/2011 01:14:42} & 55815.56 & 1200 & FIES  \\
15 & \scriptsize{12/09/2011 01:42:40} & 55816.58 & 1200 & FIES  \\
16 & \scriptsize{08/11/2011 22:55:49} & 55874.46 & 900 & HERMES  \\
17 & \scriptsize{26/10/2012 00:15:58} & 56226.52 & 1200 & HERMES \\
18 & \scriptsize{26/10/2012 23:13:34} & 56227.47 & 1200 & HERMES  \\
19 & \scriptsize{28/01/2013 19:55:47} & 56321.33 & 1200 & FIES  \\
20 & \scriptsize{29/01/2013 20:08:48} & 56322.34 & 900 &  FIES \\
21 & \scriptsize{30/01/2013 19:45:34} & 56323.32 & 900 & FIES \\
22 & \scriptsize{05/02/2013 20:40:58 }& 56329.36 & 1300 & FIES  \\
&&&&\\
\hline                  
\end{tabular}                     
\end{center}            
\end{table}             

HERMES  is a fibre-fed prism-cross-dispersed high-resolution echelle spectrograph mounted in a temperature-controlled room and fibre-fed from the Nasmyth A focal station through an atmospheric dispersion corrector (ADC). A full description of the instrument can be found in  \citet{hermes}. HERMES reaches a resolving power  of $R= 85\,000$ (or 63\,000 for the low-resolution fibre) and a spectral coverage from 377 to 900~nm, though some small gaps exist beyond 800~nm.

FIES is a cross-dispersed high-resolution echelle spectrograph, mounted in a heavily insulated building separated from and adjacent to the NOT dome,
 with a maximum resolving power of $R=67\,000$. The entire spectral range 370-730~nm is covered without gaps in a single, fixed setting \citep{fies}. In the present study, we have used the low-resolution mode with $R=25\,000$.

The HERMES data were homogeneously reduced using version 4.0 of the
HermesDRS\footnote{http://www.mercator.iac.es/instruments/hermes/hermesdrs.php}
automated data reduction pipeline offered at the telescope. A complete set of bias, flat, and arc frames obtained each night, was used to this aim. For wavelength calibration, HERMES uses a combination of a thorium-argon lamp equipped with a red-blocking filter to cut off otherwise saturated argon lines and a neon lamp for additional lines in the near infrared. The final HERMES spectra of BD~$+60^\circ\,$73 have a signal-to-noise ratio (S/N) in the $\sim80$\,--\,120 range. The HermesDRS pipeline provides wavelength-calibrated, blaze-corrected, order-merged spectra. We then used our own programs developed in IDL to normalize and correct the final spectra for
heliocentric velocity.

The FIES spectra were homogeneously reduced using the
FIEStool\footnote{http://www.not.iac.es/instruments/fies/fiestool/FIEStool.html}software in advanced mode. A complete set of bias, flat, and arc frames obtained each night, was used to this aim.  For wavelength calibration, FIES uses arc spectra of a ThAr lamp. The final FIES spectra of BD~$+60^\circ\,$73 have a S/N typically in the $\sim100$\,--\,150 range. The FIEStool pipeline provides wavelength-calibrated, blaze-corrected, order-merged spectra, and can also be used to normalise and correct the final
spectra for heliocentric velocity.
In total, we have a sample of 22 high-quality spectra obtained at different epochs spanning more than three years of observations.

\subsection{X-ray data}

The All Sky Monitor \citep[ASM;][]{levine1996} on board the {\it Rossi X-ray Timing Explorer} ({\it RXTE}) consisted of three Scanning Shadow Cameras (SSCs) mounted on a motorized rotation drive. Each SSC contained a position-sensitive proportional counter (PSPC) that viewed the sky through a slit mask covering a field of view (FOV) of $6\degr$ by $90\degr$ FWHM. Each SSC was sensitive in the energy range of approximately 1.5\,--\,12~keV, with on-axis effective areas of $\sim$10~cm$^2$, $\sim30$~cm$^2$, and $\sim23$~cm$^2$ at 2, 5, and 10~keV, respectively.

The ASM monitored the sky between January 1996 and December 2011. The ASM light curves are publicly available\footnote{http://xte.mit.edu/ASM\_lc.html} in the form of light curves covering the 1.5\,--\,12~keV energy band on two timescales: a series of 90~s exposures known as dwells, and the `one-day average' light curves, where each data point represents the one-day average of the fitted source count rates from a number (typically 5--10) of individual ASM dwells. Between 1996 January 5  (MJD 50087) and 2011 December 31 (MJD 55927), there are 82105 dwells on IGR~J00370+6122 in the ``raw'' light curve and 5407 data points in the ``one-day average'' light curve. 

The Burst Alert Telescope \citep[BAT; ][]{bart2005} on board {\it Swift}, a NASA mission launched in November 2004 \citep{ger2004}, is a coded-aperture imager with a very wide FOV of about two steradians, which operates in the 15\,--\,150~keV band. The BAT angular resolution is $22\arcmin$ (FWHM).

The BAT continually monitors the sky, with more than about 70\% of the sky observed on a daily basis. Results from this survey are publicly available\footnote {http://swift.gsfc.nasa.gov/docs/swift/results/transients/index.html} in the form of light curves covering the 15\,--\,50~keV energy band on two timescales: a single {\it Swift} pointing ($\sim$20~min) and the weighted average for each day.  This BAT daily light curve  contains 2690 data points from IGR~J00370+6122, obtained between 2005 February 15 (MJD 53416) and 2013 June 2 (MJD 56445).

\section{Data analysis \& results}
\label{data}

One of the highest quality spectra of BD~$+60^\circ\,$73 is shown in Fig.~\ref{spectrum}, where line identifications are also provided. 
The spectrum was artificially degraded to a resolving power of $R=4\,000$ for spectral classification. A large number of MK standards, observed with the same instrumentation and subjected to the same procedure, were used for comparison. The spectrum, which of much better quality that those used in \citet{reig2005}, indicates a slightly later spectral type B0.7 (based on the ratio of \ion{Si}{iii} and \ion{Si}{iv} lines, the main spectral type indicator) and somewhat higher luminosity. The best spectral type from direct comparison would be B0.7\,Ib--II, with evidence of strong nitrogen over-abundance. However, we note that the star is a very fast rotator ($v \sin i\approx135\:{\rm km}\,{\rm s}^{-1}$; see Sect.~\ref{model}), and this is known to generally lead to the assignment of lower luminosity, as the standards rotate more slowly\footnote{The number of Galactic early-B supergiants with projected rotational
velocities above $80\:{\rm km}\,{\rm s}^{-1}$ is marginal; most of them concentrate around 40--$80\:{\rm km}\,{\rm s}^{-1}$ \citep[see Fig.~14 in][]{ssimon14}.}. In view of this, we take BN0.7\,Ib.

\subsection{Distance}
\label{secdistance}

\begin{figure*}[ht!]
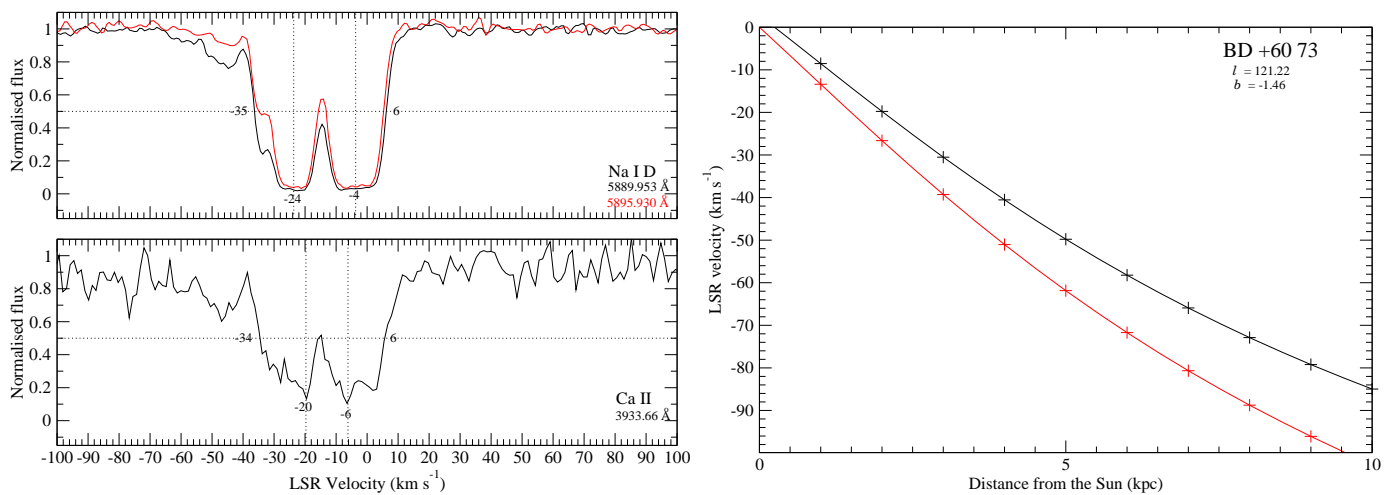

\resizebox{\columnwidth}{!}{
\includegraphics[clip, angle=0]{libdp6073_Ca.eps}}
\resizebox{\columnwidth}{!}{
\includegraphics[clip, angle=0]{d_vlsr_bd6073.eps}}
\caption{{\bf Top left: }Interstellar Na\,{\sc i} D doublet (5890\AA$\:$ full line, 5896\AA$\:$ dashed line) in the spectrum of BD~$+60^\circ\,$73. {\bf Bottom left: } The \ion{Ca}{ii} K line at 3934\AA, displaying a similar morphology, but somewhat lower S/N. {\bf Right: }Radial velocity with respect to the local standard of rest (LSR) due to galactic rotation as a function of distance, using the rotation curve of \citet[black curve]{brand93} and that of \citet[red curve]{reid09}.\label{distance} }
\end{figure*}

Traditionally, BD~$+60^\circ\,$73 has been considered a member of the Cas~OB4 association \citep{humphreys1978}. For this association \citet{humphreys1978} gives a distance modulus $DM=12.3$, while \citet{garmany} give $DM=12.2$. Photoelectric $UBV$ photometry for BD~$+60^\circ\,$73 is provided by \citet{hiltner} and \citet{haug1970}, who give essentially identical values, suggesting no variability ($V=9.65$). Therefore we can combine the optical photometry of \citet{haug1970} with the near-infrared photometry from 2MASS \citep{skru06}. By using these data as input for the $\chi^{2}$ code for parametrised modelling and characterisation of
photometry and spectroscopy {\sc chorizos}
implemented by \citet{maiz04}, we found that the extinction to the source can be reproduced well by a standard law with $R=3.1$.

If we assume an intrinsic colour $(B-V)_{0}$=$-0.20$, \citep[interpolating in the calibration of][]{fitz1970}, we obtain $E(B-V)=0.77$. The synthetic spectrum has the same intrinsic $(B-V)_{0}=-0.20$. For a standard extinction law, $A_{V}=2.39$, and then $m_V=7.26$. For a distance of $DM=12.3$, this leads to an absolute magnitude of $M_{V}=-5.0$~mag. This is a rather low value for a supergiant, though it is marginally consistent with the spectral type. Absolute magnitude calibrations for a B0.7\,Ib star would
support a value close to $M_{V}=-5.7$~mag \citep[e.g.][]{turner80,humphreys84}. However, this luminosity would place BD~$+60^\circ\,$73, at a distance of $\sim4$~kpc, which is far away from any spiral arm in this direction \citep[e.g.][]{negueruela03}.

Another way of estimating the distance to BD~$+60^\circ\,$73 is to make use of the interstellar lines in its spectrum to study the radial velocity distribution of interstellar material along its line of sight. We calculate the velocity scale with respect to the local standard of rest (LSR) by assuming that the Sun's motion with respect to the LSR corresponds to $+16.6\:{\rm km}\,{\rm s}^{-1}$ towards Galactic coordinates $l=53\degr$; $b=+25\degr$. In Fig.~\ref{distance}, we show the interstellar Na\,{\sc i} D lines; both present identical morphologies, with two well-separated components. The \ion{Ca}{ii} K line is also shown, displaying a very similar morphology. 

In the Na\,{\sc i} D lines, one of the components is centred on $-4\:{\rm km}\,{\rm s}^{-1}$, extending from low positive values to $\approx-12\:{\rm km}\,{\rm s}^{-1}$. The second component is centred on $-24\:{\rm km}\,{\rm s}^{-1}$ and has a shoulder extending to $\approx-35\:{\rm km}\,{\rm s}^{-1}$.
The \ion{K}{i}~7699\AA\ line, which is not saturated, has two narrow components, centred on $-8\:{\rm km}\,{\rm s}^{-1}$ and $-21\:{\rm km}\,{\rm s}^{-1}$. These two components are readily identified with absorbing clouds located in the Local Arm and the Perseus arm.

In Fig.~\ref{distance}, we also show the dependence of radial velocity with distance in the direction to BD~$+60^\circ\,$73 ($l=121\fdg2$, $b=-01\fdg5$) according to two widely used Galactic rotation curves. One is computed assuming circular galactic rotation and adopting the rotation curve of \citet{brand93}, with a circular rotation velocity at the position of the Sun ($d_{{\rm GC}}=8.5$~kpc) of $220\:{\rm km}\,{\rm s}^{-1}$. The other one follows the rotation curve of \citet{reid09}. Along this line of sight, all LSR velocities should be negative and increase with distance. Absorption at positive velocities must therefore be caused by clouds in the immediate vicinity of the Sun. The nearby ($l=120\fdg8$, $b=+0\fdg1$) B1\,Ia supergiant $\kappa$~Cas, located at a distance $d\sim 1$~kpc \citep{humphreys1978}, only shows the low-velocity component. Its lines have exactly the same profile on the positive side, though they are broader on the negative side and more saturated, because this object is closer to the 
Galactic plane.

The core of the high-velocity component is located around $v_{{\rm LSR}}=-24\:{\rm km}\,{\rm s}^{-1}$. This velocity corresponds to a distance $d\approx2.5$~kpc according to the rotation curve of \citet{brand93} and $\approx1.8$~kpc according to the rotation curve of \citet{reid09}. Both distances indicate that this component arises from material in the Perseus arm. There are very large deviations from circular motion in the Perseus arm  \citep{reid09}, but we can use objects with independent distance determinations to constrain the distance to the absorbing material. The nearby \ion{H}{ii} region Sh2-177 ($l=120\fdg6$, $b=-0\fdg1$) has $v_{{\rm LSR}}=-34\:{\rm km}\,{\rm s}^{-1}$, while Sh2-173 ($l=119\fdg4$, $b=-0\fdg8$) has  $v_{{\rm LSR}}=-35\:{\rm km}\,{\rm s}^{-1}$ \citep[in both cases, undisturbed gas velocities;][]{fichblitz84}. Both  \ion{H}{ii} regions have published distances compatible with the estimates for Cas OB4. More recently, \citet{russeil07} has estimated a distance $d=3.1$~kpc for Sh2-
173 ($DM=12.5$), which they also assume to be valid for Sh2-172 and Sh2-177.
We can assume that the material producing absorption in the feature centred on  $\approx-33\:{\rm km}\,{\rm s}^{-1}$ must be at the same distance. In view of this,  BD~$+60^\circ\,$73 must be at least at the same distance and most likely not much further away. 

There is some indication in all the interstellar lines of a weak component at $v_{{\rm LSR}}\approx-45\:{\rm km}\,{\rm s}^{-1}$. However, the existence of this component would not necessarily imply greater distance. Along this line of sight, there are very large deviations from circular motion, and the nearby \ion{H}{ii} region Sh2-175 ($l=120\fdg4$, $b=+2\fdg0$), with a distance estimate of $1.7$~kpc, has a velocity $v_{{\rm LSR}}=-50\:{\rm km}\,{\rm s}^{-1}$ \citep{fichblitz84}. Therefore, if this weak component at high velocities is real, it is likely to arise from foreground material associated to this \ion{H}{ii} region. In view of this, we accept a distance of 3.1~kpc for BD~$+60^\circ\,$73. This implies an absolute magnitude $M_{V}=-5.2$~mag.

\subsection{Model fit}
\label{model}

 We first applied the {\sc iacob-broad} IDL tool described in \citet{ssimon14} to estimating the projected rotational velocity of the star and the amount of macroturbulent broadening ($\Theta_{{\rm RT}}$) affecting the line profiles. 
The analysis of the \ion{Si}{iii}~4553\AA\ line resulted in the combination ($v \sin i$, $\Theta_{\rm RT}$)\,=\,(135, 67)$\:{\rm km}\,{\rm s}^{-1}$ providing the best fitting solution. While such a large macroturbulent broadening contribution is commonly found in early B supergiants, the projected
rotational speed of this star is very high compared to objects of similar spectral type \citep[e.g.][and references therein]{markova14,ssimon14}.

A quantitative spectroscopic analysis was subsequently performed by means of 
\textsc{Fastwind} (\textsc{F}ast \textsc{A}nalysis of \textsc{ST}ellar atmospheres with \textsc{WIND}s; \citealt{santo1997, puls2005}), a spherical, 
non-LTE model atmosphere code with mass loss and line-blanketing. 
An initial analysis was done following the strategy described in \cite{castro12}, based on an automatised $\chi^{2}$ fitting of synthetic \textsc{Fastwind} 
spectra including lines from \ion{H}, \ion{He}{i-ii}, \ion{Si}{ii-iv}, \ion{Mg}{ii}, \ion{C}{ii}, \ion{N}{ii-iii}, and \ion{O}{ii} to the global
spectrum between 3900 and 5100 \AA. The parameters and abundances derived were then carefully revised by using a more refined grid of \textsc{Fastwind}
models specifically constructed for this study and a {\em by eye} comparison of the observed and synthetic spectra for individual diagnostic lines
of interest (see Table \ref{lines}). In both cases the value of the wind-strength $Q$-parameter was fixed to a value $\log Q=-13.0$, because H$\alpha$ shows weak and variable emission (see Fig.~\ref{Halpha_phase}), rendering the determination of wind properties rather uncertain. The line-broadening $v \sin i$ and $\Theta_{\rm RT}$ parameters were fixed to the values indicated above during the analysis process.

The parameters obtained are summarised in  
Table~\ref{atm}. We found very good agreement (within the uncertainties) between the global-$\chi^{2}$ and the individual-line {\em by eye} solutions.
The best fit is obtained for $T_{\rm eff} =24\,000$~K and $\log g = 2.90$. The temperature value is typical of the spectral type. The low effective 
gravity fully supports the supergiant classification. 
Because of the high rotational velocity, the effective gravity has an important centrifugal contribution. When corrected for this effect, the actual 
gravity becomes  $\log g_{\rm c} = 3.00$ \citep{repolust2004}. 

\begin{table}
\caption{\label{lines} Diagnostic lines considered in the {\em by eye} individual-line analysis (see text for explanation).}
\begin{center}
\begin{tabular}{c}
\hline
\hline
\noalign{\smallskip}
H$\beta$, H$\gamma$, H$\delta$\\
\ion{He}{i}\,4471, 4387, 5875, 6678\\
\ion{Si}{iii}\,4552, 4567, 4574, \ion{Si}{iv}\,4116 \\
\ion{O}{ii}\,4661, \ion{Mg}{ii}\,4481, \ion{C}{ii}\,4267, \ion{N}{ii}\,3995 \\
\hline
\end{tabular}
\end{center}
\end{table}

\begin{table}
\caption{\label{atm} Stellar parameters derived from the spectroscopic analysis (upper panel) and calculated using the photometry of \citet{haug1970} and assuming $d=3.1$~kpc (lower panel).}
\begin{center}
\begin{tabular}{cc}
\hline
\hline
\noalign{\smallskip}
$v\sin i$ (${\rm km}\,{\rm s}^{-1}$) & 135$\pm$7\\
$\Theta_{{\rm RT}}$ (${\rm km}\,{\rm s}^{-1}$) & 67$\pm$7\\
$T_{\rm eff}$ (K) & 24\,000$\pm$1\,500\\
$\log g$ & 2.9$\pm$0.2\\
$\xi_{\rm t}$ (${\rm km}\,{\rm s}^{-1}$) & 15$\pm$5\\
$Y_{\rm He}$ & 0.25$\pm$0.10 \\
$\log Q$ & $-13.0$ (assumed) \\
\hline
\noalign{\smallskip}
$d$(kpc)&$3.1\pm0.3$\\
$M_{V}$&$-5.2\pm0.3$\\
$R_{*}$ (R$_\odot$) & $16.5\pm2.3$ \\
$\log (L_{\ast}$/$L_{\sun}$)&$4.91\pm0.16$\\
$M_{\ast}/M_{\sun}$&$10\pm5$\\
\hline
\end{tabular}
\end{center}
\end{table}

The He relative abundance ($Y_{\rm He}=N({\rm He})/N({\rm H})$) is very poorly constrained. Depending on the diagnostic line considered and the value of the
microturbulence, the derived abundance ranges from 0.15 to 0.30. Overall, the best fitting solution is obtained for $Y_{\rm He}=0.25$.
This high He abundance is indicative of a high degree of chemical evolution. The other abundances derived (C, N, O, Si,Mg) are summarised in 
Table~\ref{abundances}. We also include for comparison the set of chemical abundances for B-type stars in the solar neighbourhood derived by \citet{nieva12} and the solar abundances
from \citet{asplund09}. While Si, Mg and O abundances are compatible with the solar ones and those proposed
by \citet{nieva12} as cosmic abundance standard,  
the star appears N-enhanced and C-depleted, suggesting a fair degree of chemical evolution, 
in agreement with the He abundance. The values are similar to those of very luminous (class Ia) early-B 
supergiants \citep{crowther06}.

\begin{table}
\caption{\label{abundances} { Chemical abundances in $\log \left(\frac{N({\rm X})}{N({\rm H})}\right)+12$ resulting from the
{\sc fastwind} spectroscopic analysis of BD~$+60^\circ\,$73. We also include for comparison the set of chemical abundances
derived by \citet{nieva12} for B-type stars in the solar neighbourhood and the solar abundances from \citet{asplund09}.}}
\begin{center}
\begin{tabular}{cccc}
\hline
\hline
&&&\\
Species & BD~$+60^\circ\,$73 & B-type stars & Sun \\
\hline
\noalign{\smallskip}
Si & 7.55$\pm$0.15 & 7.50$\pm$0.05 & 7.51$\pm$0.03 \\
Mg & 7.55$\pm$0.15 & 7.56$\pm$0.05 & 7.60$\pm$0.04 \\
C & 7.75$\pm$0.15  & 8.33$\pm$0.04 & 8.43$\pm$0.05 \\
N & 8.40$\pm$0.15  & 7.79$\pm$0.04 & 7.83$\pm$0.05 \\
O & 8.75$\pm$0.15  & 8.76$\pm$0.05 & 8.69$\pm$0.05 \\
\hline
\end{tabular}
\end{center}
\end{table}

\begin{figure}
\resizebox{\columnwidth}{!}{
\includegraphics[clip, angle=0, bb = 30 50 580 420]{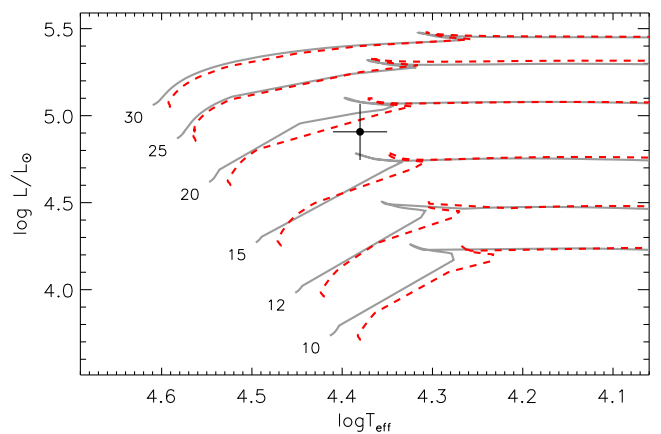}}
\caption{\label{evolutionary} Evolutionary tracks and the position of  BD~$+60^\circ\,$73 in the Hertzsprung-Russell diagram. 
The evolutionary tracks are from \citet{brott11}, for initial $v_{{\rm rot}}=0$ (grey, continuous) and $v_{{\rm rot}}=377\:{\rm km}\,{\rm s}^{-1}$ (red, dashed).\label{hr}}
\end{figure}

Stellar parameters can be derived by scaling the absolute magnitude of the synthetic spectrum to the observed $M_{V}$. 
With this method, and using the values found above for the effective temperature and the effective gravity corrected for rotation, 
we calculate the parameters listed in the bottom panel of Table~\ref{atm}. Alternatively, the mass can be estimated by placing the 
star on evolutionary tracks. In Fig.~\ref{hr}, we compare the position of BD~$+60^\circ\,$73 with the evolutionary tracks of 
\citet{brott11} for the assumed distance $d=3.1$~kpc. We show tracks with both zero and high initial rotational velocity. 
The position of BD~$+60^\circ\,$73 is compatible with the tracks for an initial mass $M_{*}$ in the $17-18\:M_{\sun}$ range close to the end 
of hydrogen core burning. Mass loss until this point is not very important, and so the present-day mass should only be a fraction of 
a solar mass lower. When the error bars are taken into account, we come to an evolutionary present-day mass of $17\pm2\:M_{\sun}$.

\subsection{Orbital solution}
\label{sec:orbit}

The radial velocity points have been obtained by cross-correlating the observed spectra  with a {\sc fastwind} synthetic spectrum 
corresponding to the parameters and abundances indicated in Tables~\ref{atm} and \ref{abundances}. The {\scriptsize IRAF} routine {\it fxcor}
was used to this aim. For the cross-correlation hydrogen and helium lines, which may be contaminated by residual wind emission in an early-type supergiant \citep[e.g.][]{vandermeer}, have been avoided, and only the region of the spectrum between 4540\AA\ and  4660\AA\ has been used. This region contains no \ion{He}{i} or \ion{H}{i} lines, but numerous metallic lines (mostly \ion{Si}{iii}, \ion{O}{ii,} and \ion{N}{ii}), which are more suitable for radial velocity determinations, because they arise in deep photospheric layers. The individual radial velocity measurements are listed in Table~\ref{OC}.

\begin{table}
\begin{center}
\caption{\label{OC}Radial velocities of BD~$+60^\circ\,$73 sorted by phase.}
\begin{tabular}{rrrr}
\hline
\hline
&&&\\
$\#$ & Phase & $v_{{\rm rad}}$ (km\,s$^{-1}$) & O-C (km\,s$^{-1}$) \\
\hline
\noalign{\smallskip}
10 &  0.0055 & $-102.1 \pm 5.5$  & $3.9$ \\
20 & 0.0390  & $-98.8 \pm 6.4$ &  $-3.5$ \\
21 & 0.1018  & $-81.7 \pm 7.5$ &  $-2.2$ \\
11 & 0.1970  & $-67.1 \pm 1.6$ &  $4.3 $\\
1 & 0.2290  & $-73.5 \pm 0.8$ &  $-3.2$ \\
2 & 0.2938  & $-72.2 \pm 0.7$ &  $-3.2$ \\
12 & 0.3259  & $-68.4 \pm 2.1$ &  $0.2 $\\
3 & 0.3525  & $-68.0 \pm 0.7$ &  $0.4 $\\
13 & 0.3869  & $-69.3  \pm 1.9 $ &  $-0.9$ \\
4 & 0.4143  & $-66.2 \pm 0.8$ &  $2.3 $\\
16 & 0.4404   & $-65.7 \pm 1.2$ &  $2.3 $\\
5 & 0.4797  & $-62.7  \pm 0.8 $ &  $6.1 $\\
22 & 0.4875  & $-73.7 \pm 0.7$ &  $-4.8$ \\
6 & 0.5460   & $-72.2 \pm 0.8$ &  $-2.7$ \\
7 & 0.6060  & $-75.2 \pm 0.8$ & $ -4.6$ \\
14 &  0.6793 & $-70.3  \pm 0.8$  & $2.1$ \\
8 &  0.7358 & $-72.0 \pm 0.7$  & $2.6$ \\
15 &  0.7444 & $-79.9 \pm 0.6$  & $-5.0$ \\
9 &  0.8705 & $-84.4 \pm 0.9$  & $0.3$ \\
17 &  0.9204 & $-92.3 \pm 1.3$  & $0.8$ \\
19 &  0.9747 & $-107.1 \pm 1.1$  & $-1.1$ \\
18 &  0.9815 & $-106.0 \pm 1.2$  & $1.0$ \\
\hline                      
\end{tabular}                       
\end{center}      
{\scriptsize Col.\,1: Number of the observation from Table~\ref{log}; Col\,2: Corresponding orbital phase according to orbital solution in Table~\ref{orbpar}; Col\,3: Residuals from the radial velocity curve (see Fig.~\ref{RadVel}) to the observational data points. }
\end{table}     

\begin{figure}
\resizebox{\columnwidth}{!}{
\includegraphics[clip, angle=270]{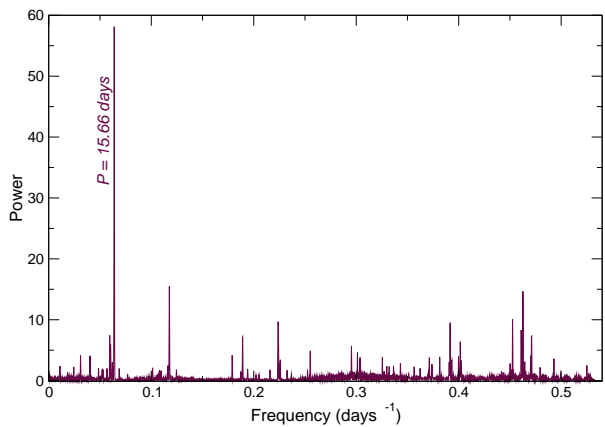}}
\caption{\label{cleanvrad} "Clean" periodogram of the radial velocity points obtained with the {\scriptsize PERIOD} package within the {\it Starlink} environment.}
\end{figure}

We searched for periodicities within the radial velocity points using different algorithms available in the {\it Starlink} package {\scriptsize PERIOD} \citep{dhillon}. All of them resulted in a highest peak at values compatible within their errors and also compatible with the periodicity found in the X-ray light curve \citep{hartog2004}. Figure~\ref{cleanvrad} shows the result obtained with the algorithm {\scriptsize CLEAN} (implemented in {\scriptsize PERIOD}), which is particularly useful for unequally spaced data. This algorithm basically deconvolves the spectral window from the discrete Fourier power spectrum producing a {\it "clean"} power spectrum. A significance calculation of the value obtained was performed using  the {\scriptsize SIG} routine of {\scriptsize PERIOD}, applying 1000 Monte Carlo simulations. The result of this significance calculation, 
representing the probability that the period is not actually equal to the value quoted, is $FAP2\sim0.7$\%. Therefore we can assume that the period of 15.66~d, which is found both from the X-ray observations \citep{hartog2004} and from the radial velocity of the optical component, is the orbital period of the binary system.

A radial velocity curve was obtained by fitting the radial velocity points using {\scriptsize SBOP}, which is an adaptation of the {\scriptsize FORTRAN~II} computer program developed to solve spectroscopic binary systems by \cite{sbop}. The fit is achieved by performing several iterations that refine a set of preliminary elements by a differential correction procedure. The initial elements of the orbit could be either supplied by the user or derived using the Russell-Wilsing method \citep{wilsing1893, russell1902, binne1960}, which is an approximation method based on a Fourier curve-fitting procedure where the only parameter that must be known is the orbital period. The differential-correction procedure can be accomplished either with the method of \cite{lehmannfilhes} re-discussed by \cite{under1966} or with the method of \cite{sterne1941} re-discussed by \cite{hiltner1962}. We fixed the period to the value obtained from the periodicity analysis, and gave a set of initial parameters as variables for a 
first iteration, assigning different weights (depending on their uncertainties) to the different radial velocity points. Using the solution of this iteration, we then utilised the Lehmann-Filh\'es method implemented in {\scriptsize SBOP} for single-lined systems. After several iterations, we obtained convergence for most orbital parameters. We then freed the period while fixing other parameters and repeated the procedure. Once we reached an accurate value of the orbital period, we fixed it to that value, freed the other parameters, and repeated again until all parameters converged.

\begin{figure}
\resizebox{\columnwidth}{!}{
\includegraphics[clip, angle=270]{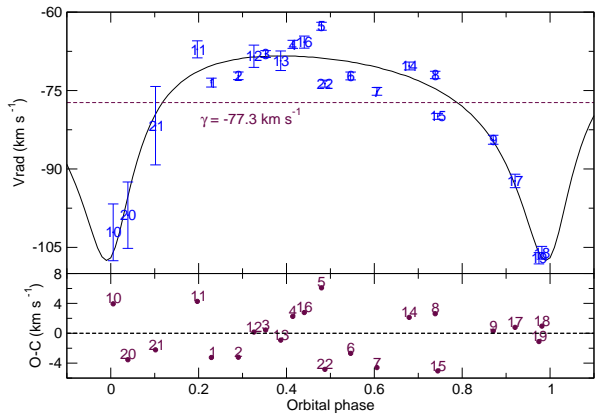}}
\caption{\label{RadVel}{\it Top:} Best-fit radial velocity curve and real data folded on the 15.661~d period derived from our data analysis. Phase 0 corresponds to periastron. {\it Bottom:} Residuals from the observed radial velocity points to best-fit radial velocity curve.}
\end{figure}

The radial velocity curve obtained is shown in Fig.~\ref{RadVel}, where the residuals from the observed data points to the theoretical curve are also shown. The parameters of this best-fit curve are shown in Table~\ref{orbpar}. The orbital period, $P_{{\rm orb}}=15.6610\pm0.0017\:{\rm d}$, is identical within the errors to the $15.665\pm0.006$~d periodicity found in the X-ray light curve by \citet{zand2007}, but slightly shorter and more precise. All parameters are well determined with the exception of the mass function, which has an error $\approx 30\%$. 

\begin{table}
\begin{center}
\caption{\label{orbpar}Orbital parameters of BD~$+60^{\circ}$73}
\begin{tabular}{cc}
\hline
\hline
\noalign{\smallskip}
$T_0$(HJD-2450000) & 5084.52$\pm$0.37\\
$P_{{\rm orb}}$ (days) & 15.6610$\pm$0.0017\\
$e$ & 0.56$\pm$0.07\\
$\gamma$ (km\,s$^{-1}$) & -77.3$\pm$1.4 \\
$K_0$ (km\,s$^{-1}$) & 19.6$\pm$1.9 \\
$\omega$ (degrees) & 194$\pm$9 \\
$a_0 \sin i$ (km) & (3.5$\pm$0.4) 10$^6$\\
$f(M)$ ($M_{\sun}$) & 0.0069$\pm$0.0023 \\
\hline
\end{tabular}
\end{center}
{\scriptsize Orbital parameters obtained by fitting the radial velocity points using Lehmann-Filh\`es method implemented in {\scriptsize SBOP}.}
\end{table}

The relatively large uncertainty is connected to the moderately large residuals in the orbital fit. In other HMXBs, such as Vela~X-1, large residuals are also seen, showing a modulation at multiples of the orbital frequency \citep{quaintrell}. This modulation led to the suggestion that the oscillations were tidally induced by the companion. Strong radial velocity excursions are also seen in the peculiar transient IGR~J11215$-$5952 \citep{lorenzo14} and in the HMXB GX~301$-$2 \citep{kaper06}, both of which have very massive and luminous counterparts. Since macroturbulence is relatively strong in BD~$+60^{\circ}$73 and a connection has been claimed between pulsations and macroturbulence broadening \citep[e.g.][]{sergio10}, it is tempting to suspect that the large residuals in the fit to the radial velocity curve may be connected to pulsations. We searched for periodicities within these residuals, finding no obvious results, though the number of measurements is probably too small to rule out the 
possibility of tidally induced pulsations completely in BD~$+60^{\circ}$73.

\subsection{X-ray light curves}

\begin{figure}
\centering
\begin{minipage}{\columnwidth}
\resizebox{\columnwidth}{!}{
\includegraphics[clip, angle=90]{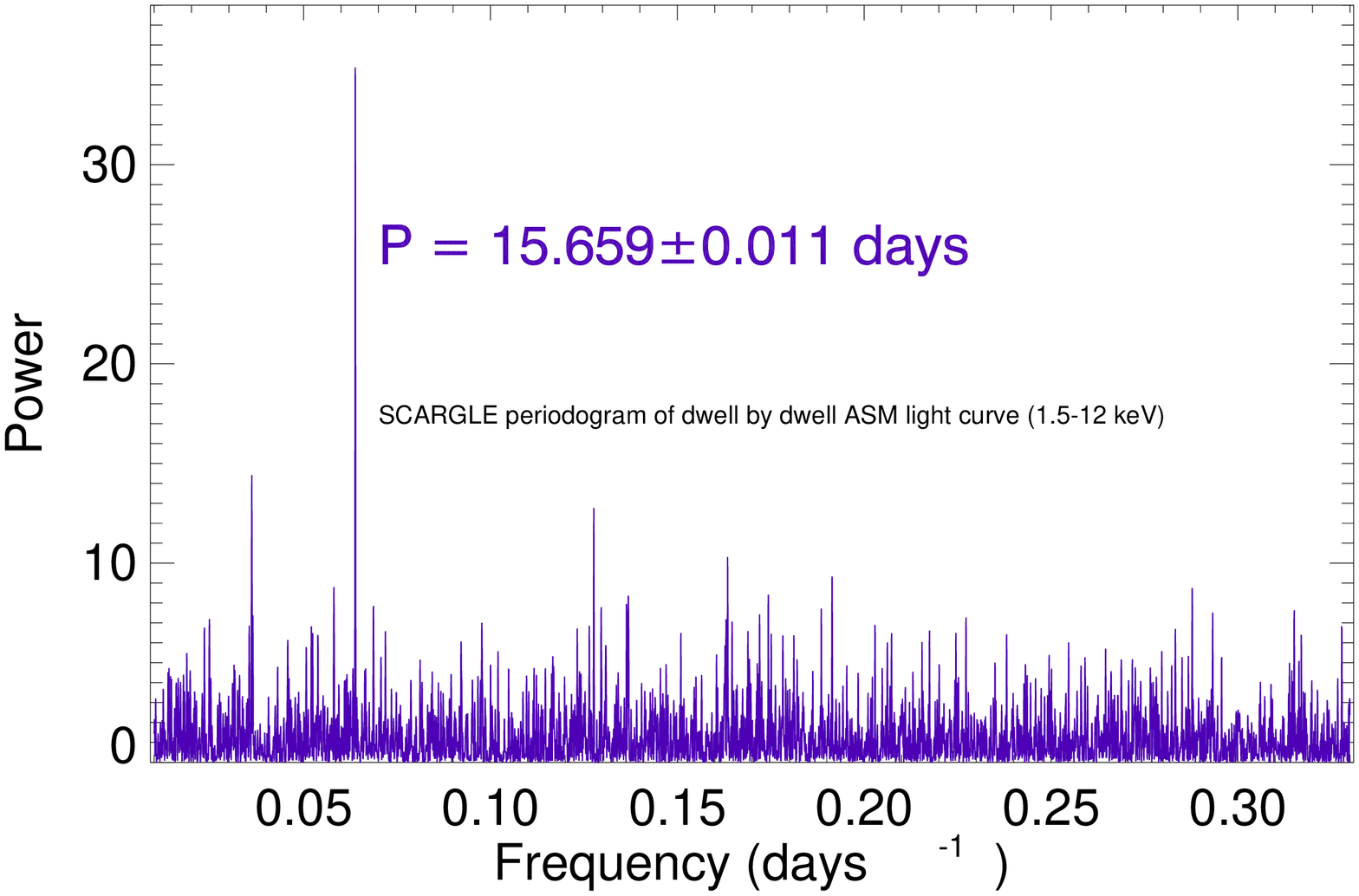}}
\end{minipage}
\begin{minipage}{\columnwidth}
\resizebox{\columnwidth}{!}{
\includegraphics[clip, angle=90]{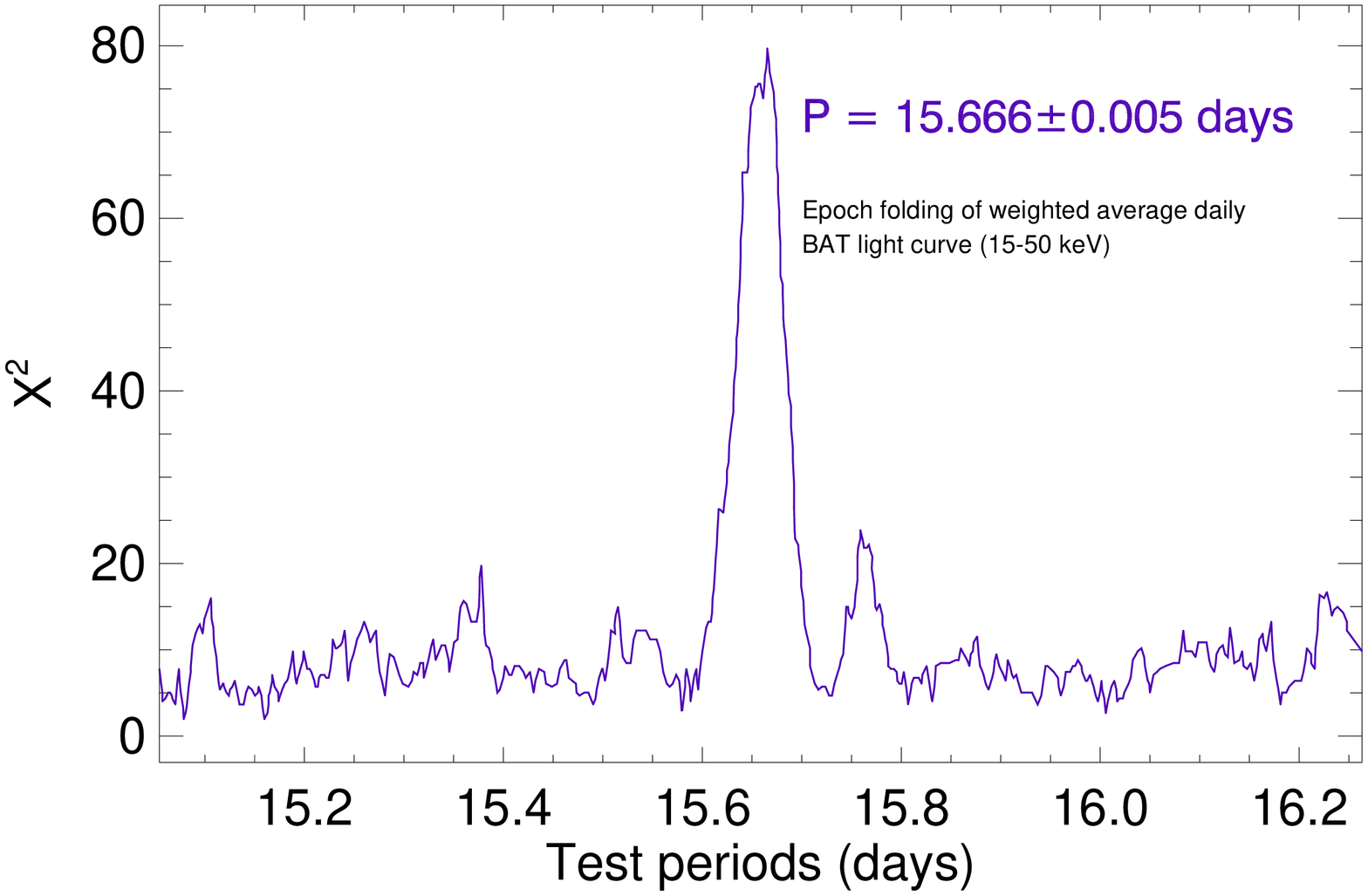}}
\end{minipage}
\caption{\label{periodsearching}  Examples of the period searching within the ASM and BAT X-ray light curves of IGR~J00370+6122. {\bf Top panel:} {\scriptsize SCARGLE} periodogram obtained with {\scriptsize PERIOD} for the `raw' (dwell by dwell) ASM  light curve. {\bf Bottom panel:} Epoch folding of the weighted-average daily BAT light curve. }
\end{figure}

The X-ray light curves from the {\it RXTE}/ASM and {\it Swift}/BAT have been searched for periodicities between 3 and 300 days, using different algorithms available within the {\it Starlink} package {\scriptsize PERIOD} and also using an epoch folding analysis implemented in the software {\scriptsize DES7} \citep[see][for details on the method implemented on this software]{larsson1996}. Analyses of the {\it RXTE}/ASM dwell-by-dwell and daily-average light curves and the BAT light curve give consistent results,  finding values for the period (see Fig.~\ref{periodsearching}) that are compatible among themselves within their respective error bars, compatible with the period found from the radial velocity points and also with the X-ray period published in \cite{zand2007} or in \cite{hartog2004}.

\begin{figure}[ht]
\resizebox{\columnwidth}{!}{
\includegraphics[clip, angle=90]{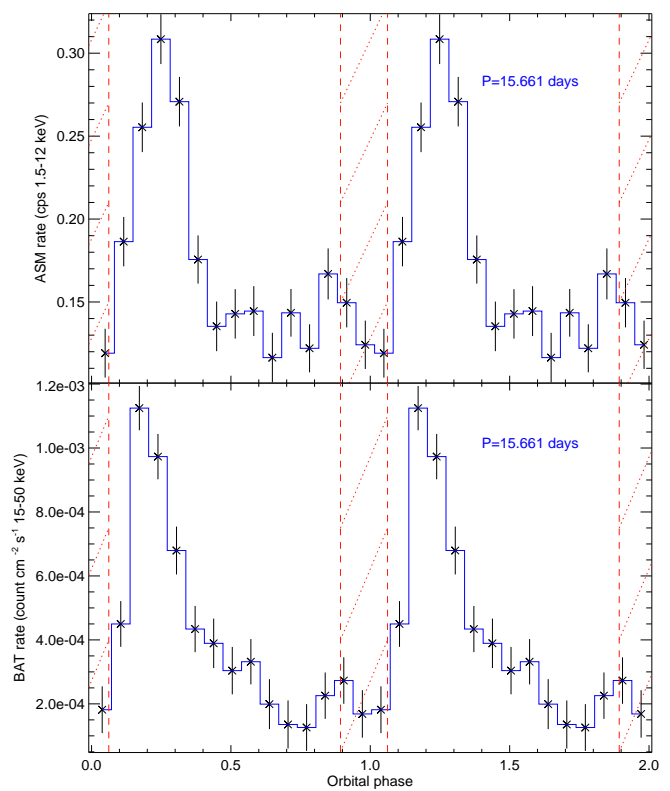}}
\caption{\label{fold} Average folded light curves in two energy ranges over the orbital period ($P_{{\rm orb}}=15.6610\pm0.0017$~days). Phase zero corresponds to the periastron passage in the {\scriptsize SBOP} orbital solution. Red dashed lines indicate the time of possible eclipse according to this solution ($\phi\epsilon[0.86-1.06]$). {\bf Top panel:} Weighted folded ASM light curve (from daily average datapoints). {\bf Bottom panel:} Weighted folded BAT light curve (from daily average datapoints).}
\end{figure}

We performed the folding of these light curves on all the periods that we had found and also on the value of $P=15.6627\pm0.0042$~d published by \citet{zand2007}. All the foldings are similar to each other in shape; the count rate has a clear maximum around $\phi=0.2$ (using the ephemeris from the radial velocity curve solution) and presents a broad flat minimum around apastron. All the foldings present some evidence of a flux drop just before the maximum, close to periastron. The highest peak is found for the folding on the orbital period derived from the radial velocity curve (which also has the smallest formal error). Therefore we interpret this as the most accurate orbital period.  Figure~\ref{fold} shows the folding of the daily average ASM light curve (top panel) and the folding of the BAT daily light curve (bottom panel) over this period. In both cases, the folding was performed with  {\scriptsize DES7}, and the individual points were weighted according to their errors.  Phase zero of these 
foldings corresponds to the periastron passage, and the figure shows the time of eclipse according to the {\scriptsize SBOP} orbital solution.

The maximum of the folded {\it RXTE}/ASM light curve is around orbital phase $\phi=0.25$ ($\sim$4~d after periastron). The shape of the light curve is very similar to the one shown in \citet{zand2007} for data taken until November 2006, in spite of the slightly different folding periods. The maximum of the {\it Swift}/BAT folded light curve is reached slightly before $\phi=0.2$, which means less than three days after periastron passage. This suggests that the maximum luminosity in the hard X-rays (energy range 15\,--\,50~keV) may occur some time before the maximum in the softer band (energy range 1.5\,--\,12~keV). The BAT light curve is very similar in shape to the ASM light curve and also quite similar to the {\it INTEGRAL}/ISGRI light curve in the 20\,--\,45~keV range shown by \citet{zand2007}, who also found some difference in shape between the soft and hard X-ray folded light curves, along with a delay between the maximum flux of the ISGRI and ASM folded light curves, with the ISGRI maximum taking place before the ASM one, which is also in accordance with our results.

The minimum of the ASM light curve is at around 0.12 cts/s. According to \citet{remillard}, the ASM has a bias of +1~mCrab and systematics make detections below 5~mCrab difficult (corresponding to $\sim 0.08$ cts/s). Therefore the source is very likely detected outside the outbursts, at a level $about ten$ times lower than  the peak luminosity, as anticipated by \citet{zand2007}. This seems to be borne out by the BAT light curve. In any case, the pointed observations with the {\it RXTE}/PCA reported by \citet{zand2007} show that these integrated fluxes represent the average of many flares superimposed on a quiescence flux at least $\sim 50$ times lower. Therefore, the folded light curves merely represent the average behaviour, in the sense that the peaks of some flares have much higher values than the peak of the folded light curves.
 
Given the spectral parameters determined by \citet{zand2007} and our value of the distance (3.1~kpc), we can estimate the X-ray luminosity in both energy ranges. Assuming a power law spectrum with an absorption column $N_{\rm H}$$=$$7\times10^{22}\:{\rm cm}^{-2}$, a photon index $\Gamma=2.3$ for the ASM energy range (1.5\,--\,12~keV), and a $\Gamma=1.84$ for the BAT energy range (15\,--\,50~keV), we determine the maximum and minimum luminosities of the average foldings as 
$L^{{\rm MIN}}_{1.5-12~{\rm keV}}$$\sim$$9.7\times10^{34}\:{\rm erg}\,{\rm s}^{-1}$, 
$L^{{\rm MAX}}_{1.5-12~{\rm keV}}$$\approx2.5\times10^{35}\:{\rm erg}\,{\rm s}^{-1}$,  $L^{MIN}_{15-50~{\rm keV}}$$\sim$$8.5\times10^{34}\:{\rm erg}\,{\rm s}^{-1}$,  
and $L^{{\rm MAX}}_{15-50~{\rm keV}}$$\approx4.7\times10^{35}\:{\rm erg}\,{\rm s}^{-1}$. \\

\subsection{Evolution of H$\alpha$}
\label{alpha}

\begin{figure}
\resizebox{\columnwidth}{!}{
\includegraphics[clip,width=7cm ]{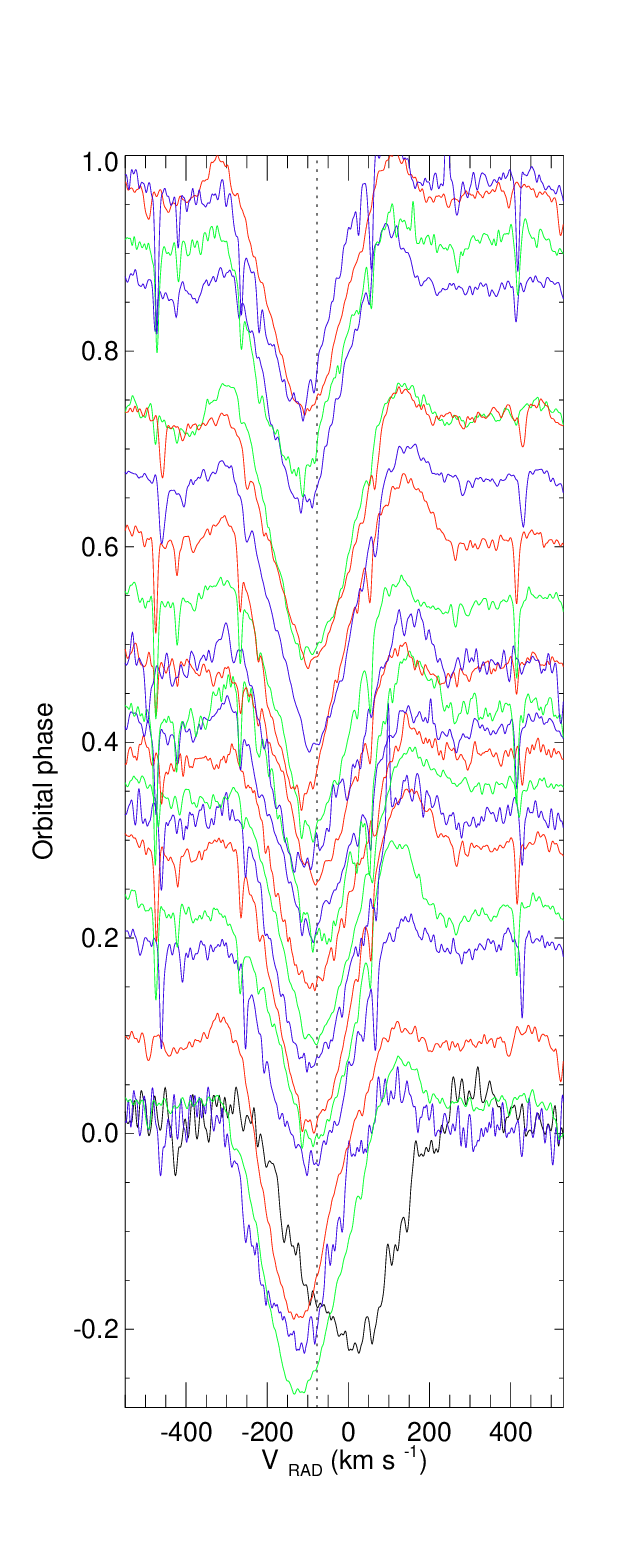}
\includegraphics[clip,width=7cm ]{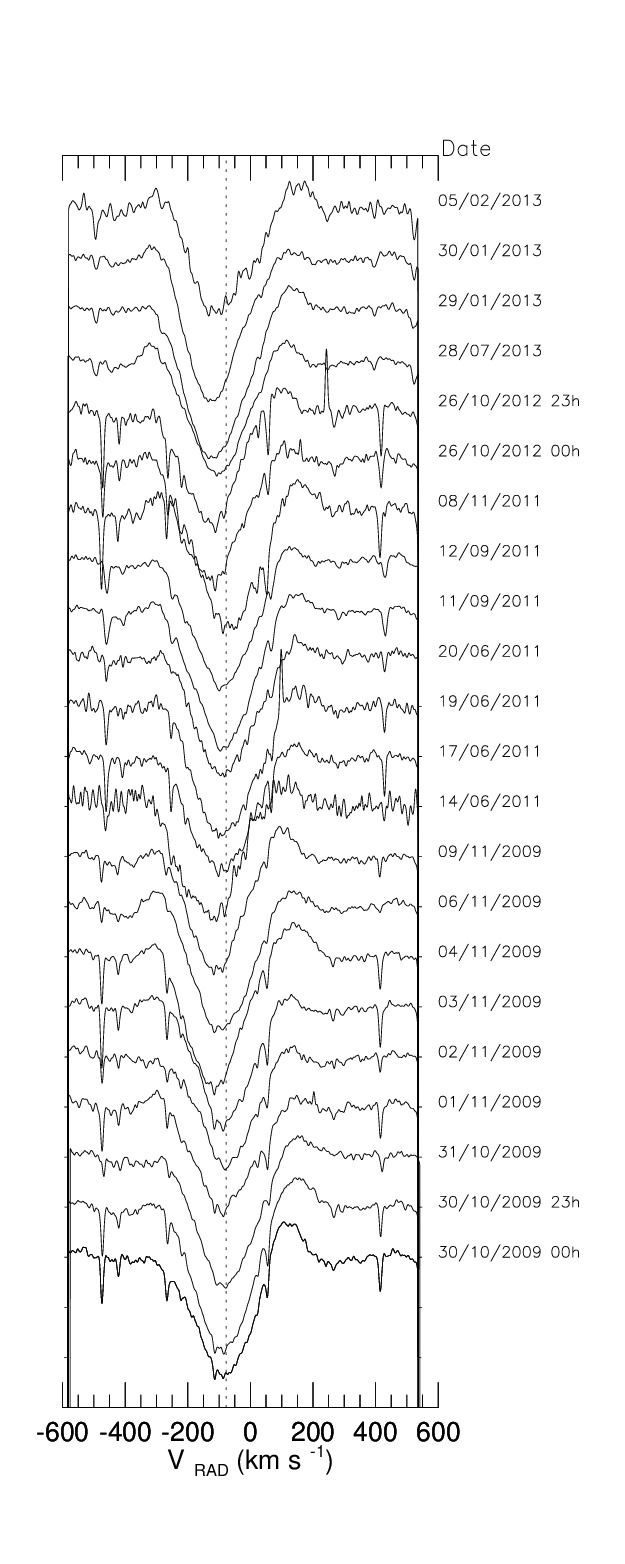}}
\caption{\label{Halpha_phase} Temporal evolution of the normalised H$\alpha$ profile, sorted by orbital phase ({\it left}) or in time sequence, with the date indicated ({\it right}).}
\end{figure}

The X-ray behaviour of IGR~J00370+6122 seems typical of wind-fed accreting sources. The shape of H$\alpha$ (Fig.~\ref{Halpha_phase}), on the other hand, is reminiscent of a Keplerian geometry and similar to those seen in shell Be stars with very small or incipient disks \citep[e.g. $o$~And;][]{clark03}. This kind of line profile, however, is also seen in OB supergiants  \citep[e.g. HD~47240;][]{morel04}. Since the neutron star must come quite close to the stellar surface at periastron, given the high eccentricity, the presence of weak emission components could be the result of mass being lost from the B supergiant and settling into some kind of disk configuration.
          
\begin{figure*}
\resizebox{\textwidth}{!}{
\includegraphics[clip,width=5.5cm,angle=90 ]{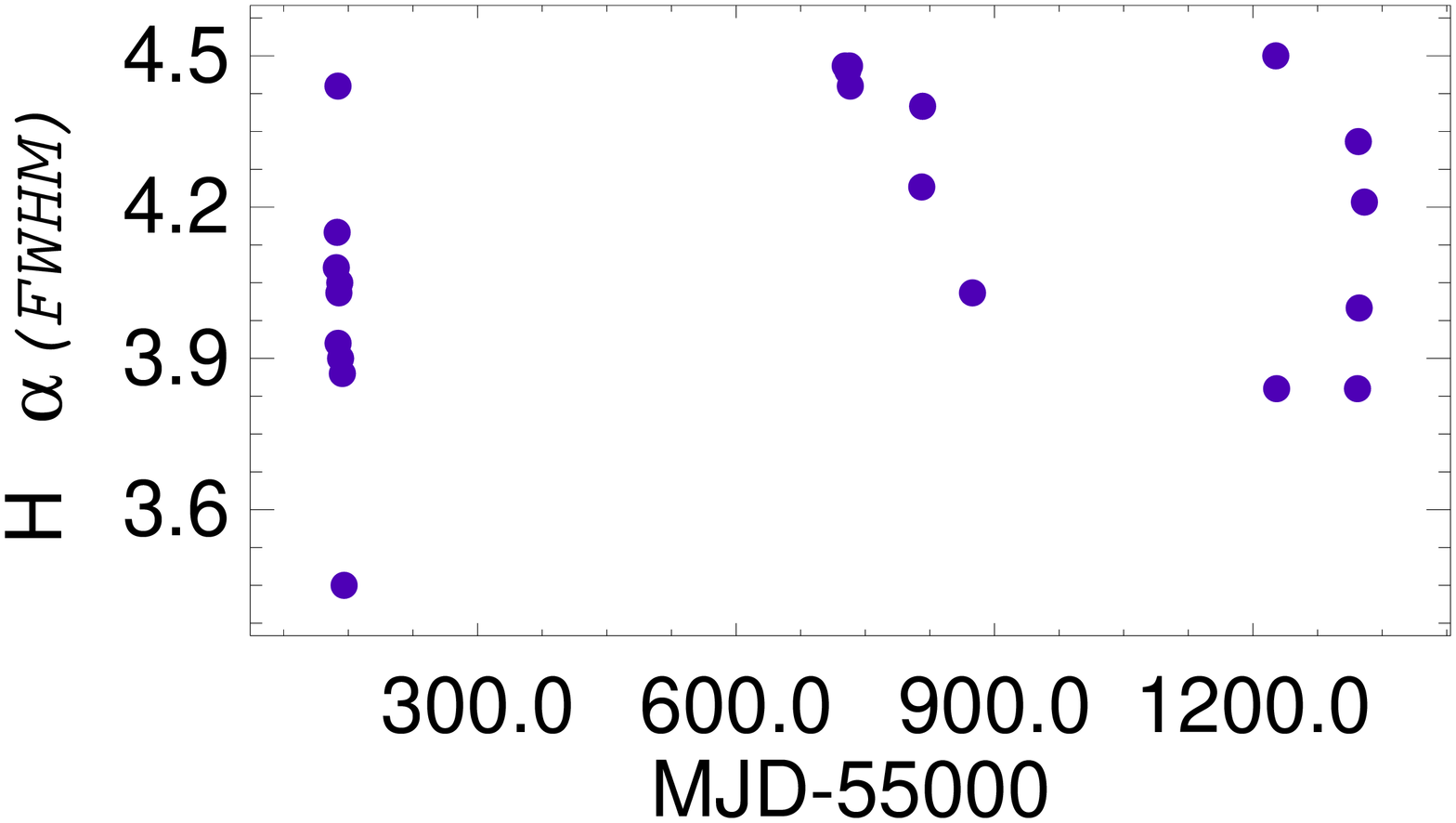}
\includegraphics[clip,width=5.5cm,angle=90 ]{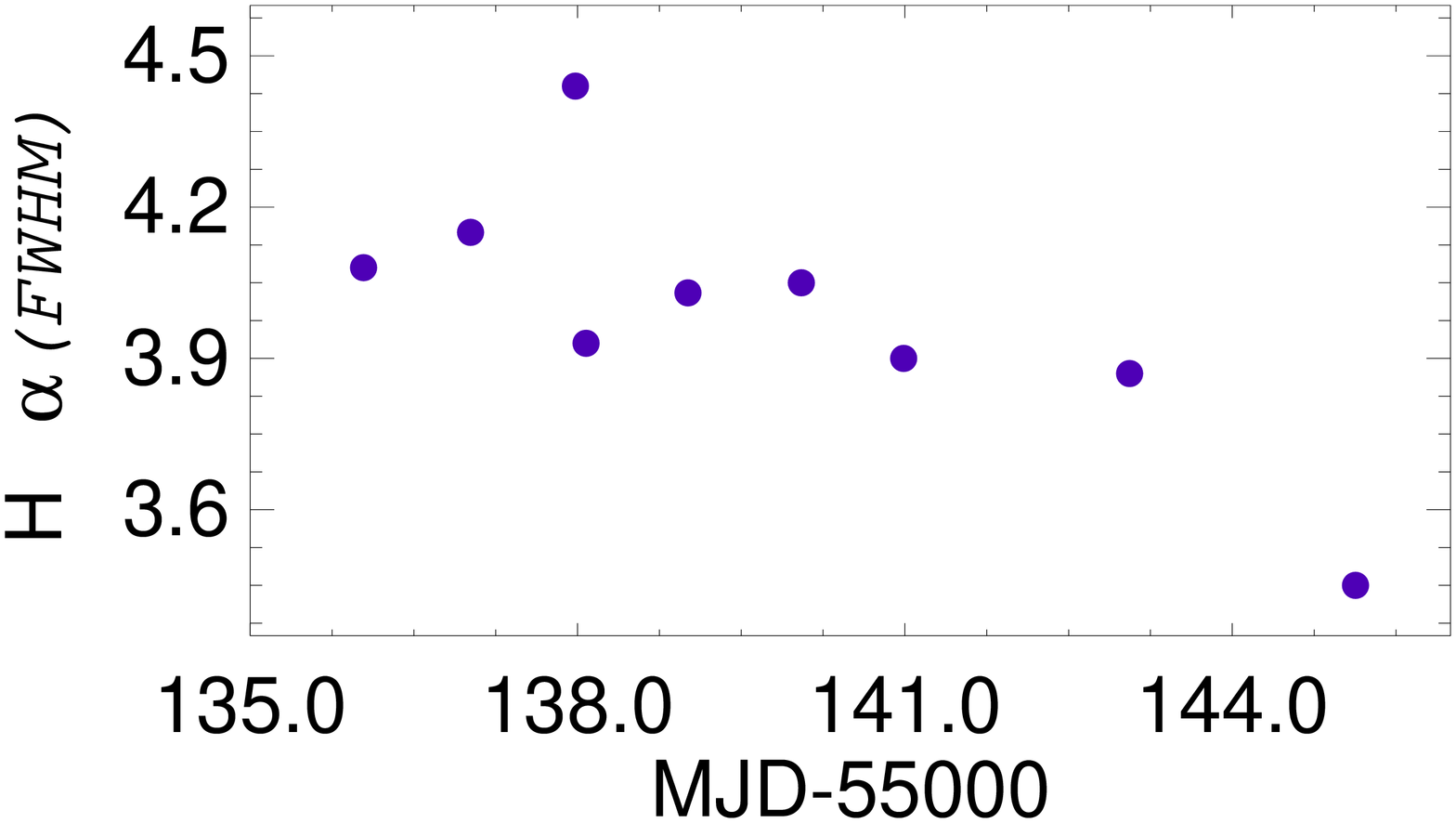}
\includegraphics[clip,width=5.5cm,angle=90 ]{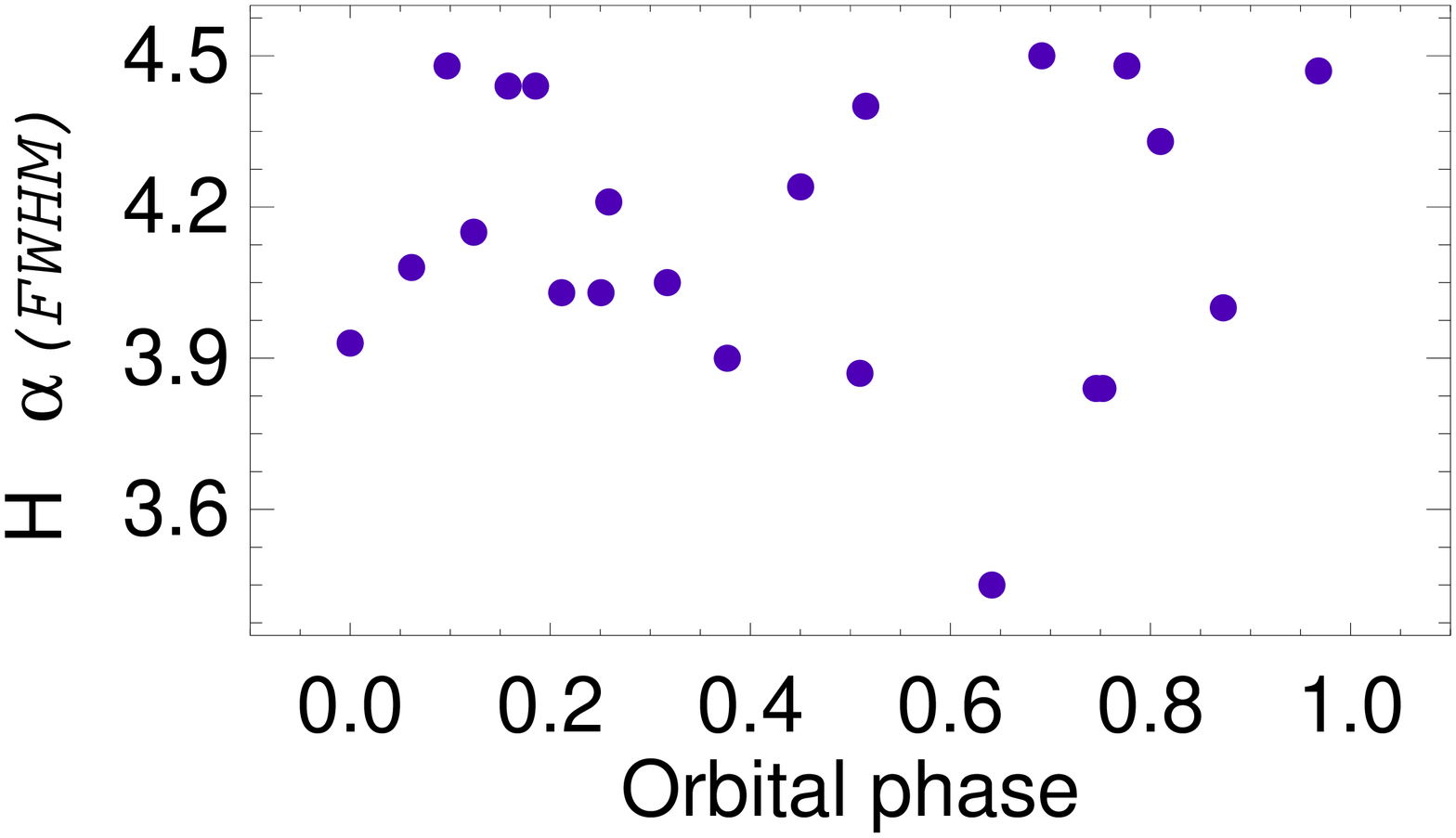}}
\caption{\label{halphafwhm} Full-width half-maximum (FWHM; in \AA) of the H$\alpha$ line. {\it Left:} Evolution with time for our whole data set. {\it Middle:} Evolution with time in the 2009 campaign with HERMES. {\it Right:} Evolution with the orbital phase of the binary system.}
\end{figure*}

To explore this possibility, Fig.~\ref{Halpha_phase} shows the evolution of H$\alpha$ ordered by orbital phase. Very similar profiles are seen at different phases, while relatively different profiles are seen at almost the same phase. This suggests that the shape of H$\alpha$ is not driven by orbital effects. To quantify this assertion, we measured the FWHM of the absorption trough by fitting a Gaussian profile to the region between 6555 and 6570\AA. The range was the same in all spectra for consistency. Figure~\ref{halphafwhm} shows the evolution of the FWHM with time and with orbital phase. The changes in the line shape seem to occur on a timescale of several days, and not to depend on the orbital phase. Therefore we conclude that the variability in H$\alpha$ is mostly due to the stellar wind, and it does not reflect episodic mass loss locked to the orbital period.

\section{Discussion} 
\label{disc}

We have determined the astrophysical parameters of BD~$+60^\circ\,$73 and obtained an orbital solution for the IGR~J00370+6122 binary system. To understand its implications for the physics of wind-accreting HMXBs, we must estimate the range of parameters allowed by our solution.

\subsection{Orbital parameters}

 The mass of the optical companion may be determined from the atmosphere parameters. In many cases, supergiants show a discrepancy between spectroscopic and evolutionary masses \citep{herrero92,herrero07}. For BD~$+60^\circ\,$73, the discrepancy is moderate. With a spectroscopic mass $M_{*}=10\pm5\:M_{\sun}$ and an evolutionary mass $M_{*}=17\pm2\:M_{\sun}$, the values are compatible at slightly above 1~$\sigma$. Since the spectroscopic value has a very strong dependence on the value of $\log g$, we favour a mass $\sim15\:M_{\sun}$, which is fully within typical values for a low-luminosity supergiant \citep[e.g.][]{markova08}. This mass is lower than those typically found in calibrations for the spectral type \citep[e.g. $22\,M_\sun$ as assumed by][]{grunhut2014} owing to its relatively low luminosity. The stellar radius, $16.5\,R_\sun$, is again smaller than is typical of the spectral type for the same reason (relatively low luminosity).

\begin{figure}
\resizebox{\columnwidth}{!}{
\includegraphics[clip,angle=90]{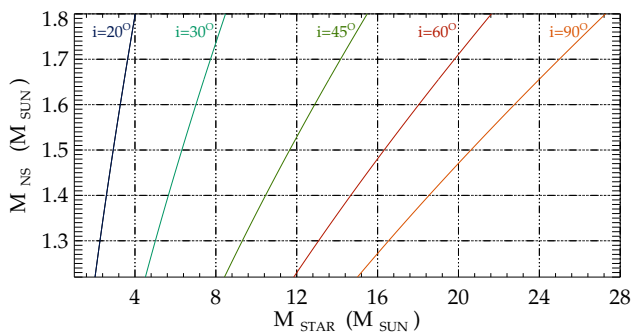}}
\caption{\label{possible} Relations between the supergiant mass and neutron star mass. Lines of constant orbital inclination constructed from the mass function are labelled.}
\end{figure}

\begin{figure}
\resizebox{\columnwidth}{!}{
\includegraphics[clip,angle=90]{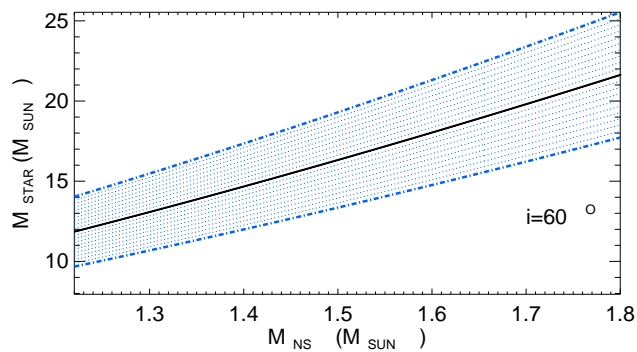}}
\caption{\label{margins} Relation between the supergiant mass and the neutron star mass for $i=60\degr$. The shaded area represents the range of values allowed within one standard deviation for our value of $f(M)$.}
\end{figure}

Unfortunately, we cannot constrain the mass of the neutron star because a pulse phase analysis does not exist. Masses of neutron stars in HMXBs range from $1.1\:M_{\sun}$ in SMC X-1 \citep{vandermeer} to $1.9\:M_{\sun}$ in Vela~X-1 \citep{quaintrell}. Since the mass function has a large relative uncertainty, we cannot impose strong constraints. Figure~\ref{possible} shows the range of possible masses for the neutron star as a function of the mass of the companion and the inclination. A companion mass $\sim15\:M_{\sun}$ rules out inclinations $\la45\degr$ (which would require neutron star masses $M_{{\rm X}}\sim2\,M_{\sun}$), and indicates an inclination not very far from $i\approx60\degr$ for a canonical $M_{{\rm X}}=1.44\,M_{\sun}$. As an illustration, Fig.~\ref{margins} shows the range of possible mass values allowed by the uncertainty in $f(M)$ when we fix $i=60\degr$. We see that for a given value of the companion mass, for instance $M_{*}=15\,M_{\sun}$, a wide range of values is allowed for the neutron star mass $M_{{\rm X}}=1.3$--1.6. In view of this, we assume a nominal value of $M_{{\rm X}}=1.44\,M_{\sun}$ for the rest of the 
paper.

\begin{figure}
\resizebox{\columnwidth}{!}{
\includegraphics[clip,angle=0]{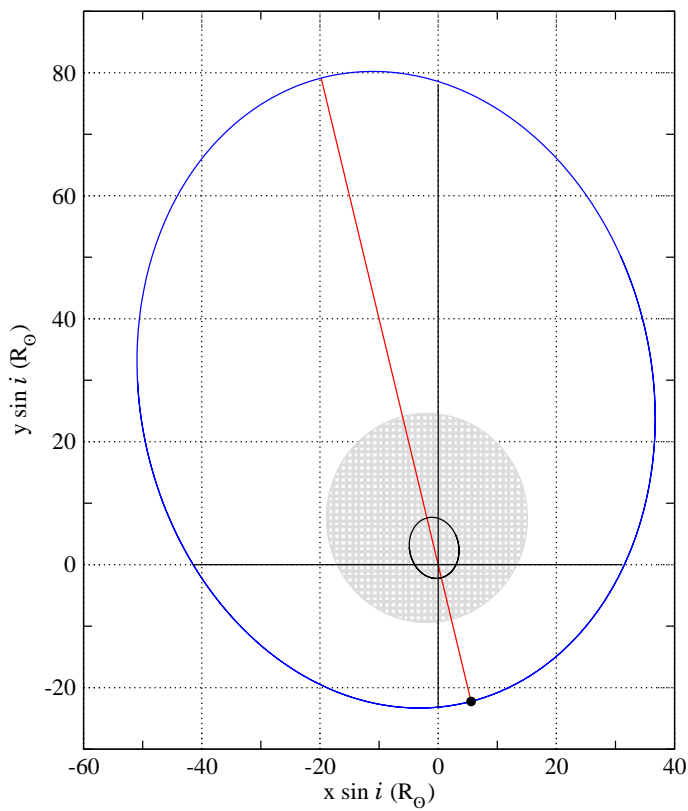}}
\caption{\label{orbita} Relative orbit of the compact object around the optical star in  IGR~J00370+6122  as seen from above the orbital plane. The position of periastron is indicated and joined to the apastron by the straight line depicting the major axis of the orbit. The shaded area signals the approximate size of the supergiant, and the small ellipse inside it represents its orbital motion around the centre of mass. The coordinates are in solar radii and represent projected distances.}
\end{figure}

A more stringent constraint on the inclination may be obtained from the likely presence of an eclipse in the folded X-ray light curves. Even though the detection cannot be statistically significant, given
the very low out-of-peak fluxes, all soft and hard X-ray light curves show evidence for a drop in flux just before the main peak. As shown in Fig.~\ref{fold}, this drop coincides with the time of possible eclipse according to our orbital solution, therefore we consider the existence of an eclipse likely. If this is actually the case, the inclination of the system must be $\ga60\degr$. In particular, for the canonical  $M_{{\rm X}}=1.44\,M_{\sun}$, and our values of $M_{{*}}=15\,M_{\sun}$ and  $R_{{*}}=17\,R_{\sun}$, the inclination must be $i\ga71\degr$. This inclination value agrees with $i\sim72\degr$ proposed by \cite{grunhut2014}, even though they assume a much larger radius ($R_{*}=35\,R_\sun$) than we derive ($R_{*}=16.5\,R_\sun$). Since we cannot constrain the duration of the eclipse, we cannot constrain the inclination any further, but values as high as $i=75\degr$ are easily compatible with the masses discussed here (see Fig.~\ref{possible}). With these masses and the eccentricity 
found in Sect.~\ref{sec:orbit}, the periastron distance is $r=36\:R_{\sun}$, i.e. $2.1\:R_{*}$. The orbit corresponding to these parameters is displayed in Fig.~\ref{orbita}.

Comparing our orbital solution with the solution found by \cite{grunhut2014}, all the parameters are compatible with each other within errors (mass function, systemic velocity, semi-amplitude, etc.) except for the time of periastron passage. This time in \cite{grunhut2014} differs by $\sim0.1$ phases (according to our ephemeris) from our value. The two values, however, are consistent within errors at the 2 $\sigma$ level, which represents good agreement if we take the $\sim133$ orbital cycles into account between the two dates. The values of the eccentricity are compatible with each other, confirming the high eccentricity. \cite{grunhut2014} have fixed the orbital period value to that of \cite{zand2007}, while we calculated an orbital period through a periodogram of the radial velocity measurements and the X-ray light curves and refined the value via the fitting of the orbital solution with {\small SBOP}. Therefore we expect our value of the orbital period to be more accurate than that of \cite{zand2007}. In conclusion, \cite{grunhut2014} find a totally compatible orbital solution using a completely independent data set. We can thus consider the orbital parameters of this HMXB firmly determined.

\subsection{Evolutionary context}

The stellar properties of BD~$+60^\circ\,$73, and in particular its high level of chemical evolution, are rather unusual. The star displays a very high $v \sin i$ for a supergiant. In addition, both the He abundance and the N/C ratio are typical of stars displaying a high degree of processing, such as B\,Ia supergiants \citep[e.g.][]{crowther06}. It is possible to relate the high level of chemical evolution to the high rotational velocity, assuming that it is the result of mass transfer from the progenitor of the neutron star prior to its supernova explosion. In classical models for the formation of SGXBs, the original primary transfers mass to the secondary, leading to its ``rejuvenation'' \citep{podsiadlowski92, wellstein01, langer12}. If mass transfer is conservative, the mass of the original secondary can grow until it becomes rather more massive than the primary initially was \citep[e.g.][]{wellstein99}. After the primary explodes as a supernova, the original secondary evolves to the supergiant phase, perhaps overluminous for its mass \citep{dray07}, developing a strong wind that feeds the neutron star.

For a limited range of initial parameters, \citet{pols94} found that mass transfer could lead to a reversal of the supernova order. The original secondary would receive the whole hydrogen-rich envelope of the primary, becoming much more massive, and then fill its Roche lobe, leading to reverse mass transfer and probably a common-envelope phase. The explosion of the original secondary would occur while the original primary is still some sort of He star. The observational characteristics of such a He star are not known. 

As explained in Sect.~\ref{model}, the He abundance of BD~$+60^\circ\,$73 is not well constrained, but is high. The most likely value is  $Y_{\rm He}=0.25$, with a permitted range between 0.15 and $\sim0.3$. The He abundances of B supergiant are generally assumed to be close to the solar value, $Y_{\rm He}=0.10$ \citep[e.g.][]{markova08}. For very luminous (luminosity class Ia) supergiants, \citet{crowther06} assumed $Y_{\rm He}=0.20$, indicative of some chemical evolution, since they found N/C ratios similar to the one found for BD~$+60^\circ\,$73. For the blue hypergiants $\zeta^{1}$ Sco and HD 190603, which are expected to be in a more advanced stage of evolution, \citet{clark12} find $Y_{\rm He}=0.20$, while \citet{kaper06} found $Y_{\rm He}=0.29$ for the  B1\,Ia$^{+}$ donor in the SGXB GX~301$-$2.

Therefore BD~$+60^\circ\,$73 may be the B supergiant with the second highest He abundance measured. Together with its relatively high N/C ratio, very high rotational velocity, and perhaps mild underluminosity (see Sect.~\ref{secdistance}), this evidence of chemical evolution may indicate a non-standard evolutionary channel for IGR~J00370+6122. However, given the lack of a theoretical evolutionary model that can reproduce these characteristics, we do not try to infer the formation history of the system and stick to the standard scenario. 

\begin{figure}
\resizebox{\columnwidth}{!}{
\includegraphics[clip,angle=90]{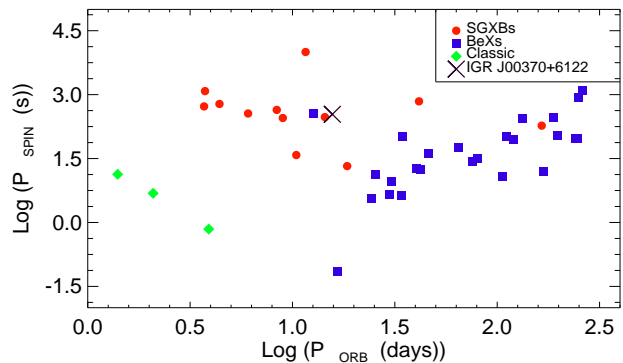}}
\caption{\label{corbet} The $P_{{\rm spin}}/P_{{\rm orb}}$ (Corbet's) diagram for a large sample of HMXBs. ``Classical'' HMXBs are fed through localised Roche-lobe overflow and have very short spin periods. Be/X-ray binaries show a moderately strong correlation between their orbital and spin periods \citep{corbet84}. Wind-fed systems have long spin periods, uncorrelated to their orbital period. The position of IGR~J00370+6122 is marked by a cross.}
\end{figure}

Within this evolutionary scenario, a major observational constraint is given by the pulse period of the neutron star. \citet{zand2007} observed IGR~J00370+6122 on four occasions with the {\it RXTE}/PCA, always close to the predicted time of maximum X-ray flux, according to the  {\it RXTE}/ASM light curve. The X-ray flux detected was very variable, with evidence of strong flaring, leading \citet{zand2007} to conclude that the ASM folded light curves simply represent averages of many flares. During the strongest of these flares, which took place on MJD 53566, reaching $L_{\rm X}(3-60\,keV)\sim3.2\times10^36\,{\rm erg}\,{\rm s}^{-1}$ \citep[][;value scaled to our distance]{zand2007}, the flux was strongly modulated. Timing analysis revealed a very probable pulsation at $P_{{\rm spin}}=346\pm6$~s that cannot be considered certain due to the brevity of the flare \citep{zand2007}. The combination of flaring behaviour and long spin period is typical of a wind-accreting system, as expected for a supergiant companion. Figure~\ref{corbet} shows the $P_{{\rm spin}}/P_{{\rm orb}}$ diagram, where different kinds of HMXBs occupy different positions according to its accretion behaviour \citep{corbet86}.

The position of IGR~J00370+6122 in Fig.~\ref{corbet} is typical of wind-fed supergiant systems, giving further support to the putative spin period. According to the classical interpretation \citep[e.g.][]{corbet84, waterskerkwijk}, the spin periods of Be/X-ray binaries are equilibrium periods for the magnetized neutron stars. The equilibrium is a consequence of spin-up episodes during accretion and spin-down episodes when accretion is inhibited by centrifugal effects \citep[see also][]{stella86}. In the case of wind-fed systems, the observed spin periods are believed to be the equilibrium values reached when the donor was still on the main sequence, because there is no effective way for a new equilibrium to be achieved when the source is a supergiant  \citep{waterskerkwijk, liu11}. In this scenario, detection of a pulse period typical of wind accretors in IGR~J00370+6122 has two immediate consequences:

\begin{itemize}
\item  IGR~J00370+6122 is not a young system. The supernova explosion of the progenitor to the neutron star took place when BD~$+60^\circ\,$73  was still close to the main sequence, since the pulsar had time to slow down from its presumably very short initial spin period to its current value.

\item In spite of the fast rotation measured in BD~$+60^{\circ}\,$73, the system did not go through a Be/X-ray binary phase before becoming a supergiant X-ray binary. \citet{liu11}  describe scenarios in which a Be/X-ray binary may turn into an SFXT as the donor star evolves off the main sequence, but the spin period of the neutron star is not expected to slow down again. The SFXT IGR~J18483$-$0311, with a putative $P_{{\rm spin}}=21$~s, was considered to be the best candidate to have followed this evolutionary path, but this spin period has recently been called into question \citep{ducci13}. We note that no Be/X-rays binaries are known with an orbital period as short as  IGR~J00370+6122. Only the peculiar LMC transient A\,0538$-$66 has a similar $P_{{\rm orb}}=16.7$~d, but this object does not seem to behave as a classical Be/X-ray binary, and the putative $P_{{\rm spin}}=0.069$~s, detected only once, sets it apart from all Be/X-ray binaries.

\end{itemize}

That IGR~J00370+6122 is not a young system indicates that it might have had time to circularise. SGXBs with orbital periods smaller than 8~d have negligible eccentricities. Vela X-1, with $P_{{\rm orb}}=8.9$~d and a B0.5\,Iab companion, has a low eccentricity of $e=0.09$, but 4U~1907+09, with $P_{{\rm orb}}=8.4$~d, and a late-O supergiant companion has $e=0.28^{+0.10}_{-0.14}$  \citep{zand98}. EXO 1722$-$363, with an early B companion and  $P_{{\rm orb}}=9.7$~d has an eccentricity less than $e=0.2$ and perhaps negligible \citep{thompson07}. This variety suggests that the birth of neutron stars results in a measurable eccentricity in (almost) all systems, but tidal circularisation reduces it in systems with small orbits.

Observational data suggest that a binary containing stars with an outer radiation zone will suffer substantial tidal effects when the ratio between the semi-major axis and the stellar radius is smaller than $\sim4$ \citep{north03}. For typical values of mass and radius for an OB star orbited by a neutron star, this corresponds to orbital periods shorter than $\sim10$~d. Three SGXBs with orbital periods longer than 10~d show measurable, but moderate eccentricity. With $P_{{\rm orb}}=10.4$~d, OAO~1657$-$415 has $e=0.11$ \citep{barnstedt}. With $P_{{\rm orb}}=11.6$~d, 2S~0114+650 has $e=0.18\pm0.05$ \citep{grundstrom}. Finally, 1E~1145.1$-$6141, with an orbital period $P_{{\rm orb}}=14.4$~d, has $e=0.20\pm0.03$ \citep{raychakrabarty}. The slightly longer orbital period of IGR~J00370+6122, together with its much higher eccentricity, $e=0.56\pm0.07$, supports the idea that the system will not have time to circularise during the lifetime of the optical
companion. The wide-orbit SGXB GX~301$-$2, with a B1\,Ia$^{+}$ donor and a $P_{{\rm orb}} =46.5$~d, has slightly lower eccentricity, $e=0.46$ \citep{kaper06}. Only the peculiar system IGR~J11215$-$5952, with $P_{{\rm orb}}=164.6$~d, is likely to have higher eccentricity \citep{lorenzo14}.

The systemic velocity of BD~$+60^{\circ}\,$73 is $v_{{\rm hel}}=-77\:{\rm km}\,{\rm s}^{-1}$, corresponding to $v_{{\rm LSR}}\approx-70\:{\rm km}\,{\rm s}^{-1}$. This systemic velocity is perfectly compatible with the $-81\pm3{\rm km}\,{\rm s}^{-1}$ addopted by \cite{grunhut2014}, who obtained different values of this parameter for their four different data sets. Our value is indeed equal to their $\gamma_3$ value \citep[see Table~3 of][]{grunhut2014}. The few members of Cas~OB4 with published velocities seem to cluster around $v_{{\rm hel}}\approx-50\:{\rm km}\,{\rm s}^{-1}$ \citep{humphreys1978}. Therefore the recoil velocity of the system after the supernova explosion was $\approx-25\:{\rm km}\,{\rm s}^{-1}$ in the radial direction. The tangential component was not very important either, because it has no significant proper motions \citep[$-1.66\pm1.04$~mas~yr$^{-1}$; $0.51\pm1.05$~mas~yr$^{-1}$;][]{van2007}. These values agree with the average of other members of Cas~OB4 (e.g. HD~1544 with $-1.68\pm0.82$~mas~yr$^{-1}$; $-1.3\pm0.6$~mas~yr$^{-1}$ or BD~$+63^{\circ}$70 with $-2.26\pm0.93$~mas~yr$^{-1}$; $-0.9\pm0.74$~mas~yr$^{-1}$, which have the smaller formal errors in the {\it Hipparchos} catalogue; \citealt{van2007}). Therefore the supernova explosion that imparted a rather high eccentricity to 
the system does not seem to have given it a very high recoil velocity. With this peculiar radial velocity,  BD~$+60^{\circ}\,$73 may have travelled 2.5~pc in 1~Myr, and therefore cannot be very far away from its natal location.

\subsection{The zoo of wind-fed X-ray binaries}

The spectral classification of BD~$+60^{\circ}\,$73, at the very limit of the Ib luminosity class, and its low absolute magnitude suggest that it is the least luminous of all mass donors in well-studied wind-fed supergiant binaries, either SGXBs or SFXTs. The relatively weak wind, combined with the rather wide orbit, results in a low X-ray luminosity, with an average peak luminosity $L_{\rm X}$$\sim$$10^{35}$~erg~s$^{-1}$, below the average X-ray luminosity of persistent SGXBs, but above the average X-luminosity of SFXTs in quiescence. With most of the X-ray luminosity
being released in a small fraction of time, its flaring behaviour is reminiscent of SFXTs, and indeed could be considered similar to those of intermediate SFXTs with long orbital periods, such as SAX~J1818.6$-$1703 \citep{zurita09}. IGR~J18483$-$0311, which has a similar orbital period $P_{{\rm orb}}=18.5$~d, displays an {\it INTEGRAL}/ISGRI 20\,--\,40~keV light curve \citep{sguera07} that is very similar in shape to that of IGR~J00370+6122 \citep{zand2007}, even if the peak luminosity is much higher, at least five times brighter for a source at approximately the same distance \citep{rahoui08}. The maximum X-ray luminosity reported to date is that of the flare detected by \cite{zand2007} with the PCA in MJD~53566, reaching $L_{\rm X}(3-60\,keV)$$\sim$$3.2\times10^{36}$~erg~s$^{-1}$ \citep[][;value scaled to our distance]{zand2007}. Compaing this value to the minimum X-ray luminosity of the folded X-ray light curves (Fig.~\ref{fold}), we find a dynamical range for the X-ray luminosity of IGR~J00370$+$6122 of $\sim10^2$, which is below the canonical value to define an SFXT \citep[$>10^3$; e.g.,][]{walter2007}. Therefore although the X-ray light curve of IGR~J00370$+$6122 is dominated by flaring behaviour (contrary to the behaviour of persistent SGXBs), this flaring does not achieve the dynamical range of SFXTs.

These ``intermediate'' systems are supposed to represent a link between sources where the X-ray emission is completely dominated by flaring (the SFXTs) and the persistent (and brighter) canonical wind-fed SGXBs. Monitoring of IGR~J18483$-$0311 with {\it Swift} over a whole orbital period revealed a dynamical range $>1\,200$ \citep{romano10}, well above the threshold used by \citet{walter2007} to classify an object as an SFXT. \citet{romano10} suggested that the X-ray light curve of IGR~J18483$-$0311 could be explained by an eccentricity $\sim0.4$. Our orbital solution for IGR~J00370+6122, together with the orbital solution of \cite{grunhut2014} that supports ours, reveals very similar orbital parameters for this system.

We know that the SFXT behaviour cannot be primarily related to the orbital size or eccentricity. SFXTs with small orbits are known, such as IGR~J16479$-$4514 \citep{jain09} or IGR~17544$-$2619 \citep{djclark09}, both with periods $<5$~d. On the other hand, IGR~J16465$-$4507 has a long orbital period, $P_{{\rm orb}}=30.2$~d, and a moderate dynamical range that make it more like canonical SGXBs \citep{parola10,djclark10}. Given the high eccentricity of IGR~J00370+6122, the neutron star must come as close to the donor star at periastron as in a persistent system like 1E~1145.1$-$6141, which is much brighter in the X-rays. The SFXT behaviour is also unlikely to be primarily related to the spin period of the neutron star, since SFXTs and SGXBs seem to have similar spin periods\footnote{However, we note that the published spin periods for IGR~J18483$-$0311 and IGR~17544$-$2619 have been recently called into question \citep{ducci13,drave14}. Since the spin period of IGR~J00370+6122 cannot be considered to be completely certain, the only SFXT left with an observed spin period is the intermediate system IGR~J16465$-$4507, with  $P_{{\rm spin}}=227$~s \citep{walter07}.}. 

This lack of a clear, immediate cause and the likely connection of X-ray flaring with accretion of high-density clumps of material in the wind \citep{rampy2009,bozzo11} has led many authors to speculate on the possibility that clump accretion is the primary cause of SFXT behaviour \citep{zand2005,walter2007}.  
However, recent theoretical considerations \citep[e.g.][]{oskinova12} and observations  \citep[e.g.][]{sidoli13b} strongly suggest that wind clumpiness is not the only explanation for the flaring behaviour, leading to the widespread idea that some ``gating'' mechanism mediates the accretion
onto the neutron star. Proposed mechanisms include a magnetic barrier that prevents direct accretion \citep[subsonic propeller regime;][]{doroshenko11} and a switch in the polar beam configuration due to a variation in
the optical depth of the accretion column \citep{shakura13}. The recently proposed ``accumulation mechanism''  \citep{drave14}, which may be a natural consequence of quasi-spherical accretion and the changes in accretion configuration described by \citet{shakura13}, seems to be an excellent candidate for this gating process  because it removes the need for the accreting objects in al SFXTs to be magnetars \citep[e.g.,][]{bozzo2008,grunhut2014}.

In addition, recent observations of high variability and off states in the canonical SGXBs 4U~1907+09 \citep{sahiner12,doroshenko12} and IGR~J16418$-$4532 \citep{drave13} help to blur the difference between persistent and transient wind-accreting X-ray binaries. The  properties of IGR~J00370+6122 make it intermediate between classical persistent SGXBs and ``intermediate'' SFXTs, adding one more point to this continuum. Moreover, as the donor star in IGR~J00370+6122 evolves, it will increase its luminosity, turning into a later-type, more luminous supergiant, similar to the counterparts of 2S~0114+65 or 1E~1145.1$-$6141. We can therefore expect  IGR~J00370+6122 to evolve into a typical, persistent SGXB, showing that a given source may move across the parameter space and pass through both the SFXT and SGXB phases.

\section{Conclusions}

We have presented a detailed spectral analysis of BD~$+60^{\circ}\,$73, the optical counterpart to IGR~J00370+6122, derived its astrophysical parameters, and described its orbital motion. We classified BD~$+60^{\circ}\,$73 as a BN0.7\,Ib star just at the lower limit in luminosity for a supergiant and find $T_{\rm eff} =24\,000$~K and $\log g = 2.90$. At a distance of 3.1~kpc, the star may be slightly underluminous for the spectral type and effective gravity found. However, the absolute magnitude derived, $M_{V}=-5.2$, is within the margins allowed by the error bars in $\log g$ and the uncertainty in spectral type allowed by the very fast rotational velocity of the star, $v \sin i\approx135\:{\rm km}\,{\rm s}^{-1}$.  The spectroscopic mass $M_{*}=10\pm5\:M_{\sun}$ is compatible with the evolutionary mass $M_{*}=17\pm2\:M_{\sun}$, leading to an assumed value of $M_{*}=15\:M_{\sun}$.

Abundance analysis reveals a high degree of chemical processing, with a N/C ratio and a He abundance typical of highly evolved stars, such as blue hypergiants. Together with the high rotational velocity, this level of chemical evolution may be an indication of a non-standard evolutionary channel or perhaps of substantial accretion of processed material from the progenitor of the neutron star.

The orbital solution reveals high eccentricity ($e=0.56\pm0.07$) and suggests a periastron distance  $r\approx2\:R_{*}$. The X-ray emission is strongly peaked around orbital phase $\phi=0.2$, though the observations are consistent with some level of X-ray activity happening at all orbital phases.
With these parameters, IGR~J00370+6122 is the most eccentric HMXB with a supergiant donor to date. Its X-ray light curve has a very similar shape to that of the ``intermediate'' SFXT IGR~J18483$-$0311, but somewhat lower average luminosity. Though its whole X-ray flux apparently consists of flares, with a very low quiescence level, the integrated flux over many orbits seems to show only a moderate contrast between the peak, centred at orbital phase $\phi=0.2$, and other phases. In this sense, its behaviour could also be interpreted as a more extreme version of the light curves of persistent SGXBs. The relatively wide orbit, high eccentricity, and low luminosity of the companion (implying only a moderate mass loss) contribute to explaining this low X-ray flux. The future evolution of the mass donor, which must expand and become more luminous (thus presenting a higher mass-loss rate), will very likely turn IGR~J00370+6122 into a persistent SGXB.

The low average X-ray luminosity, together with the flaring nature of the high-energy emission, prevents confirmation of a possible eclipse, for which we find hints in all long-term folded light curves slightly before periastron (located at $\phi=0$). The possibility of the eclipse is reinforced by the fact that the mass function suggests a high inclination $i\ga60^{\circ}$. Observations with high-sensitivity missions are needed to confirm or reject this hypothesis, which would imply an inclination $i\sim70\degr$.

Aside from the inclination angle, the astrophysical parameters of IGR~J00370+6122 are well constrained. Its properties support  the idea that the difference between persistent SGXBs and SFXTs are unlikely to be explained by a single, simple factor, such as higher eccentricity or wind clumping, or even a combination of both, and, together with other recent findings, strongly point towards the existence of some gating mechanism that controls the accretion flow on to the neutron stars in wind-fed X-ray systems.\\

\begin{acknowledgements}

We thank the referee, Dr. Roland Walter, for constructive criticism that led to substantial improvement of the paper.
We thank Prof. Norbert Langer for fruitful discussions on binary formation. We also thank Dr. Carlos Gonz\'alez-Fern\'andez for participating in some HERMES observations.
The work of A.G. has been supported by the Spanish MICINN under FPI Fellowship BES-2009-014217 associated to grant AYA2008-06166-C03-03. This research is partially supported by the Spanish Mineco under grants AYA2010-21697-C05-04/05 and AYA2012-39364-C02-01/02. SS-D acknowledges financial support from the Spanish Ministry of Economy and Competitiveness (MINECO) under the 2011 Severo Ochoa Program MINECO SEV-2011-0187. The work was partly based on observations made with the Mercator Telescope, operated on the island of La Palma by the Flemish Community, at the Spanish Observatorio del Roque de los Muchachos of the Instituto de Astrof\'isica de Canarias and partly on observations made with the Nordic Optical Telescope, operated on the island of La Palma jointly by Denmark, Finland, Iceland, Norway, and Sweden in the Spanish Observatorio del Roque de los Muchachos of the Instituto de Astrof\'isica de Canarias. We kindly acknowledge the staff at the Nordic Optical and Mercator Telescopes for their professional competence and always useful help.

\end{acknowledgements}

\bibliographystyle{aa}
\bibliography{bib}

\end{document}